\documentclass[
preprint,
 amsmath,amssymb,
 aps,
 pre,
]{revtex4-2}
\usepackage{graphicx}
\usepackage{dcolumn}
\usepackage{bm}
\usepackage[colorlinks=true, allcolors=blue]{hyperref}

\usepackage{amsmath}
\usepackage{amsfonts}
\usepackage{url}
\usepackage{xcolor}
\usepackage{setspace}
\usepackage[caption=false]{subfig}
\usepackage{newtxtext}
\usepackage{newtxmath}

\newcommand{\FN}{f\mathcal{N}}

\newcommand{\noisecoeff}{\xi}
\newcommand{\exkurt}{{\gamma_2}}
\newcommand{\GammaShape}{{\tilde{\alpha}}}
\newcommand{\GammaRate}{{\tilde{\beta}}}

\newcommand{\uFluid}[0]{\boldsymbol{u}}
\newcommand{\bcdot}[0]{\boldsymbol{\cdot}}
\newcommand{\bnabla}[0]{\boldsymbol{\nabla}}

\newcommand{\ESpecific}[0]{\mathcal{E}}

\newcommand{\StressTensor}[0]{{\boldsymbol{\tau}}}
\newcommand{\HeatFlux}[0]{{\boldsymbol{Q}}}

\newcommand{\lrangle}[1]{{{\langle #1 }\rangle}}          
\newcommand{\lrangleS}[1]{{{\langle #1 }\rangle_\mathrm{s}}}       
\newcommand{\lrangleN}[1]{{{\langle #1 }\rangle_{(N)}}}   
\newcommand{\lrangleDT}[1]{{{\langle #1 }\rangle_{\Delta t}}}     
\newcommand{\lrangleM}[1]{{{\langle #1 }\rangle_\mathrm{m}}}       

\newcommand{\half}{{\frac12}}
\newcommand{\kB}{k_\mathrm{B}}
\newcommand{\Tbar}{\overline{T}}

\begin{document}

\title[]{Variance and Higher-Order Moments in Computational Fluctuating Hydrodynamics with Multiplicative Noise}

\author{Satinderpreet Singh}
\affiliation{Department of Applied Mathematics, University of California, Merced, California 95343, USA}

\author{Changho Kim}
\email{ckim103@ucmerced.edu}
\affiliation{Department of Applied Mathematics, University of California, Merced, California 95343, USA}

\author{Ishan Srivastava}
\affiliation{Center for Computational Sciences and Engineering, Lawrence Berkeley National Laboratory, Berkeley, California 94720, USA}

\author{Andrew J. Nonaka}
\affiliation{Center for Computational Sciences and Engineering, Lawrence Berkeley National Laboratory, Berkeley, California 94720, USA}

\author{John B. Bell}
\affiliation{Center for Computational Sciences and Engineering, Lawrence Berkeley National Laboratory, Berkeley, California 94720, USA}

\author{Alejandro L. Garcia}
\affiliation{Department of Physics and Astronomy, San Jose State University, San Jose, California 95192, USA}

\date{\today}

\begin{abstract}
Fluctuating hydrodynamics (FHD) models mesoscopic fluid systems by augmenting classical conservation laws with stochastic flux terms governed by the fluctuation-dissipation theorem. 
While traditional validation of computational FHD (CFHD) algorithms focuses primarily on second-order central moments of equilibrium fluctuations, accurately resolving higher-order moments—such as skewness and kurtosis—is essential for evaluating non-linear functions of hydrodynamic fields. 
In this work, we analyze finite-volume CFHD discretizations to evaluate second-, third-, and fourth-order central moments of thermodynamic fluctuations.
Using the one-dimensional stochastic heat equation, we demonstrate that incorporating multiplicative noise is necessary to recover the expected gamma distribution for temperature in the canonical ensemble, whereas additive noise models yield a Gaussian distribution that fails to capture non-zero skewness and excess kurtosis. 
We also derive analytical expressions for inherent modeling errors that persist in the limit of vanishing time step for a variety of boundary conditions. 
Furthermore, we introduce an improved formulation for Dirichlet boundary conditions that eliminates large boundary-adjacent errors in the third central moment present in traditional boundary treatments. 
We also examine higher-order temporal integration schemes, demonstrating that Stratonovich-based integrators, such as Heun’s predictor-corrector method, can reduce these modeling errors. 
Finally, numerical simulations of the full compressible, stochastic Navier--Stokes equations confirm the generality and practical applicability of the findings obtained from the stochastic heat equation.
\end{abstract}

\maketitle

\section{Introduction}

Fluctuating hydrodynamics (FHD) extends classical continuum mechanics to the mesoscale, where thermal fluctuations play an important role in the overall dynamics. 
While macroscopic fluid mechanics relies on deterministic conservation laws, the hydrodynamic fields for systems at the micro- and nanoscale are inherently stochastic. 
To capture these effects without the computational expense of molecular dynamics (MD), FHD models mesoscopic dynamics using a generalization of the Navier--Stokes equations introduced by Landau and Lifshitz~\cite{Landau_59,Zarate_07}. 
This approach leads formally to stochastic partial differential equations (SPDEs) that augment the standard hydrodynamic equations with stochastic flux terms that model spontaneous thermal fluctuations. 
The magnitudes of these noise terms are determined by the fluctuation-dissipation theorem, which ensures that the stochastic forcing is balanced by physical dissipation to maintain the correct equilibrium Gibbs--Boltzmann distribution.

Although early work showed promise~\cite{Garcia_JSP_87,Bell_JCP_07,Bell_PRE_07}, numerically solving FHD equations presents unique challenges because standard discretization schemes often fail to preserve fluctuation-dissipation balance at the discrete level. 
An approach to treating FHD equations is to introduce finite volume discretizations that are specifically designed to satisfy a discrete version of fluctuation dissipation balance~\cite{Donev_CAMCOS_10,balboa2012staggered}.  
This recasts the FHD equations as a system of stochastic ODEs with multiplicative noise. 
Explicit stochastic Runge--Kutta time integration methods have been developed to discretize the equations in time while preserving the structure of the fluxes. 

For the most part, the validation of CFHD schemes has focused on second-order central moments of fluctuations.
For example, in the case of temperature fluctuations, the variance $\lrangle{ \delta T^2(\mathbf{r}) }$ measured in a CFHD calculation can be compared with a known theoretical result, usually for systems at thermodynamic equilibrium.
Since fluctuation dissipation balance is built into the spatial discretization, the measured deviations from the expected theoretical values have been primarily ascribed to numerical truncation error due to the temporal discretization.
However, recasting the FHD equations as SODEs with multiplicative noise also introduces ``modeling errors'' that do not vanish in the limit $\Delta t \rightarrow 0$. One of the objectives of this study is to quantitatively assess both theoretically and numerically these modeling errors. 

The accuracy of fluctuations for higher-order moments (3rd, 4th, etc.) of hydrodynamic variables for CFHD has largely gone unexplored.\footnote{We note that Cornalba and Fischer~\cite{cornalba2023dean} prove a weak convergence estimate for higher moments of the Dean--Kawasaki equation.}
However, accurately resolving higher-order moments can play an important role for some applications. For example, suppose that we have a non-linear function of temperature, $g(T)$, which is evaluated in our simulation; some good examples would be a nucleation rate or the equilibrium constant for a reaction. 
At steady state, we have $\lrangle{ g } = \int dT g(T) \mathcal{P}(T)$ where $\mathcal{P}$ is the stationary distribution for temperature.
The Taylor expansion of $g(T)$ about the average temperature $\Tbar$ shows that
\begin{equation}
    \lrangle{g} = g(\Tbar) 
    + \frac12 \left. \frac{d^2 g}{dT^2} \right|_{T=\Tbar} \lrangle{\delta T^2} 
    + \frac16 \left. \frac{d^3 g}{dT^3} \right|_{T=\Tbar} \lrangle{\delta T^3} 
    + \ldots \, ,
\end{equation}
where $\delta T = T - \Tbar$.
Clearly, if $g(T)$ is asymmetric about $\Tbar$ then the value of $\lrangle{ g(T) }$ obtained in the simulation will be more accurate if the central moments of temperature are correct.

In the present study, we consider one of the simplest fluctuating hydrodynamic equations, the stochastic heat equation (SHE)~\cite{Garcia_JSP_87,garcia2025introCFHD}.
The one-dimensional SHE is
\begin{equation}
    \partial_t T(x,t) = \kappa \partial_x^2 T + \noisecoeff \partial_x\, (T \tilde{Z}) \, ,
\end{equation}
where $\kappa = \lambda/\rho c_V$ is the thermal diffusivity. 
Here, $\rho$ is density, $c_V$ is the specific heat capacity and $\lambda$ is the thermal conductivity, which are taken as constants.
By the fluctuation-dissipation theorem, the noise coefficient is $\noisecoeff = \sqrt{2 \kB \lambda}/\rho c_V$ where $\kB$ is the Boltzmann constant.
The noise variance is
\begin{equation}
    \lrangle{\tilde{Z}(x,t) \tilde{Z}(x',t')} 
    = \frac{1}{\mathcal{A}} \delta (x - x') \delta (t - t') \, ,
\end{equation}
where $\mathcal{A}$ is the cross-sectional area of the system.

Obtaining high accuracy in CFHD calculations is important for many reasons, particularly when we consider non-equilibrium scenarios. 
For example, consider the one-dimensional stochastic heat equation with a linear temperature gradient.
For Dirichlet boundary conditions at $x= 0$ and $\ell$, the correlation of temperature fluctuations is~\cite{Garcia_JSP_87}:
\begin{equation}
    \langle \delta T(x) \delta T(x') \rangle 
    = \frac{\kB \overline{T}(x)^2}{\rho c_V \mathcal{A}} \delta (x-x') 
    + \frac{\kB (\partial_x \overline{T})^2}{\rho c_V \mathcal{A} \ell} \times 
        \begin{cases} 
            x(\ell-x'), & \text{if } x < x', \\ 
            x'(\ell-x), & \text{otherwise}.
        \end{cases}
\label{eq:TcorrNEQ}
\end{equation}
The first term on the RHS is the local equilibrium variance and the second is a weak, long-range correlation of the temperature fluctuations.
Even in the extreme case where $\overline{T} \approx |\ell \partial_x \overline{T}|$, the ratio of the non-equilibrium to equilibrium terms is typically only a fraction of a percent.
Although these non-equilibrium correlations are typically weak, they can produce significant macroscopic phenomena, such as the giant fluctuation effect~\cite{VailatiGiglio1997,DonevBellDeLaFuenteGarcia2011,BaaskeBatallerBraibantiCarpinetiCerbinoCroccoloDonevKöhlerOrtizDeZárateVailati2016}.
More recently, the work of Bussoletti, \textit{et al.}~\cite{bussoletti2026non} indicates that away from equilibrium the three-point skewness of concentration fluctuations in a binary fluid depends on the concentration gradient due to a nonlinear mode coupling of concentration and velocity fluctuations.

In this paper, we analyze finite-volume CFHD schemes for the stochastic heat equation to obtain expressions for the second, third, and fourth central moments of temperature.
This analysis reveals (and numerical simulations confirm) that these central moments are accurately obtained by these numerical schemes with multiplicative noise.
Our treatment covers both closed systems with Dirichlet boundary conditions and isolated systems with either periodic or homogeneous Neumann boundary conditions.  For isolated systems, we explicitly account for finite-size corrections due to energy conservation.
From this analysis, we are also able to derive a new treatment of Dirichlet boundary conditions that dramatically improves resolution of higher-order moments compared to the traditional Dirichlet boundary treatment.
Furthermore, we obtain expressions for the small modeling errors resulting from the discretization of the stochastic partial differential equation.  These errors persist even in the limit of vanishing time step.
Finally, we demonstrate the generality of our observations with measurements of central moments in a CFHD computation of the full stochastic, compressible Navier--Stokes equations.

\section{Temperature Fluctuations}\label{sec:MomentsTheory}

Our interest here is to determine whether numerical solutions of the stochastic heat equation accurately yield the correct central moments for temperature fluctuations.
We first consider the case where the probability distribution of states is given by the canonical ensemble, that is, for a system in contact with a heat reservoir~\cite{callen1985}. 
For the stochastic heat equation, this would correspond to Dirichlet boundary conditions at fixed temperature.
The canonical ensemble also gives the limiting case of a small subvolume (grid cell) in an infinitely large system that acts as the ``heat bath'' holding the subvolume at an equilibrium temperature.
As shown in Section~\ref{sec:CanonicalEnsemble}, the temperature fluctuations in the canonical ensemble are given by the gamma distribution.

We are also interested in cases where a system is isolated (e.g., periodic boundary conditions), in which case the central moments for the temperature fluctuations differ from those of the gamma distribution.
Specifically, for the stochastic heat equation, the fluctuations of temperature in a subvolume of a finite system depend on the total system size due to conservation of energy.
Rather than analyzing the general case, as given by the microcanonical ensemble, in Section~\ref{sec:ConservationCorrection} we formulate a simple correction to the central moments that accounts for energy conservation.
As we shall see, this conservation equation, which represents a physical effect, and the modeling error terms provide an accurate prediction of numerical simulations.

\subsection{Canonical Ensemble}\label{sec:CanonicalEnsemble}

From Callen~\cite{callen1985}, the $\mathrm{n}^{th}$ central moments of the energy fluctuations are obtained from the probability distribution for the canonical ensemble as
\begin{equation}
    \lrangle{\delta E^n} 
    = \lrangle{(E - U)^n} 
    = \sum_j^\mathrm{states} \left(E_j-U\right)^n \exp\left(\beta(F-E_j)\right) \, ,
\end{equation}
where $\beta = 1/\kB T$, $E$ is the energy of a state, $U = \lrangle{E}$ is the equilibrium value of the energy, and $F$ is the free energy for the canonical ensemble (i.e., the Helmholtz free energy).
Callen gives the recursion relation for the fluctuations in the moments,
\begin{equation}
    \lrangle{\delta E^{n+1}} 
    = -\frac{\partial}{\partial\beta} \lrangle{\delta E^n} 
    - n \lrangle{\delta E^{n-1}} \frac{\partial U}{\partial\beta} \, .
\end{equation}
The second moment is
\begin{equation}
    \lrangle{\delta E^2} = \rho \kB c_V \Tbar^2 V \, ,
\end{equation}
where $V$ is the volume of the system.
Using the recursion relation, for constant $c_V$ the third and fourth moments are
\begin{equation}
    \lrangle{\delta E^3} = 2 \rho \kB^2 c_V \Tbar^3 V \, ,
    \quad
    \lrangle{\delta E^4} = 3 \rho c_V \kB^2 (\rho c_V V + 2\kB) \Tbar^4 V \, .
\end{equation}
These results are for the canonical ensemble, that is, a system in contact with a heat bath.
For other cases (e.g., isolated systems), the moments are the same as above for subsystems of volume $V$ with a correction that is $O(V/V_\Sigma)$ where $V_\Sigma$ is the total volume of the system (see Section~\ref{sec:ConservationCorrection}). 

Our interest is in temperature fluctuations so using the above and the relation $E = \rho c_V V T$ we have
\begin{equation}
    \lrangle{\delta T^2} = \frac{\kB \Tbar^2}{\rho c_V V} \, ,
    \quad
    \lrangle{\delta T^3} = \frac{2 \kB^2 \Tbar^3}{\rho^2 c_V^2 V^2} \, ,
    \quad
    \lrangle{\delta T^4} = \frac{3 \kB^2 (\rho c_V V + 2\kB) \Tbar^4}{\rho^3 c_V^3 V^3} \, .
\label{eq:T_moments_orig}
\end{equation}
It is useful to write the specific heat capacity as $c_V = \frac12 f \kB / m$ where $f$ is the number of classical degrees of freedom and $m$ is the atomic mass.
For example, for a classical solid, by the law of Dulong and Petit, $c_V = 3 \kB/m$ for $f=6$ degrees of freedom (3 translational plus 3 vibrational).
Using this relationship, we obtain $\rho c_V V = \frac12 f \kB \mathcal{N}$, where $\mathcal{N}$ is the number of molecules in volume $V$, and
\begin{equation}
    \lrangle{\delta T^2} = \frac{2 \Tbar^2}{f\mathcal{N}} \, ,
    \quad
    \lrangle{\delta T^3} = \frac{8 \Tbar^3}{f^2 \mathcal{N}^2} \, ,
    \quad
    \lrangle{\delta T^4} 
    = \frac{\left(12 + \frac{48}{f\mathcal{N}} \right) \Tbar^4}{f^2 \mathcal{N}^2} 
    = \left(3 + \frac{12}{f\mathcal{N}}\right) \lrangle{\delta T^2}^2 \, .
\label{eq:T_moments_fN}
\end{equation}
From the central moments, we can compute the coefficient of variation, skew, and kurtosis
\begin{equation}
\begin{split}
    \mathrm{CV}[T] 
    &= \frac{\sqrt{\lrangle{\delta T^2}}}{\lrangle{T}} 
    = \sqrt{\frac{2}{f\mathcal{N}}} \, ,
    \quad
    \mathrm{Skew}[T] 
    = \frac{\lrangle{\delta T^3}}{\lrangle{\delta T^2}^{3/2}} 
    = \sqrt{\frac{8}{f\mathcal{N}}} \, ,
    \\
    \mathrm{Kurt}[T] 
    &= \frac{\lrangle{\delta T^4}}{\lrangle{\delta T^2}^2} 
    = 3 \left(1 + \frac{4}{f\mathcal{N}} \right) \, .
\label{eq:T_SkewKurt}
\end{split}
\end{equation}
Note that the distribution is leptokurtic ($\mathrm{Kurt}[T] > 3$), which means it has ``heavy'' tails.

As proven in Appendix~\ref{app:GammaDist}, in the canonical ensemble the probability distribution function for temperature is the gamma distribution
\begin{equation}
    \mathcal{P}_\gamma(T) 
    = \frac{\GammaRate^\GammaShape}{\Gamma(\GammaShape)} T^{\GammaShape-1} e^{-\GammaRate T} \, ,
\end{equation}
with shape parameter $\GammaShape = \frac12 f\mathcal{N}$ and rate parameter $\GammaRate = \GammaShape / \Tbar$.
This is not surprising since for particle velocities that are Maxwell--Boltzmann distributed, their kinetic energy is gamma distributed.

\subsection{Conservation Correction}\label{sec:ConservationCorrection}

The results above for the canonical ensemble are no longer exact for subvolumes of a finite, isolated system. 
A system is isolated when all boundaries are periodic or if they satisfy a homogeneous Neumann boundary condition.  
The total energy in an isolated system is conserved and, as such, the distribution of temperature fluctuations must be consistent with this conservation property.

Anticipating our finite-volume discretization of the stochastic heat equation, consider an isolated system divided into $N$ cells, each of volume $V$. 
For any conserved quantity, the sum of the fluctuations must be zero so that the number of independent fluctuations is $N-1$ not $N$. 
We expect this correction to be small when $N$ is very large. 
However, when the number of cells is small, say $O(10)$, the distribution $\mathcal{P}(T)$ deviates noticeably from the gamma distribution.

To quantify the correction due to conservation, we consider the ansatz that a correction to the variance and quadratic correlations can be written as
\begin{equation}
    \lrangleN{\delta T_i \delta T_j} = a_1 \delta_{ij} + a_2 \, ,
\label{eq:ansatz}
\end{equation}
where the subscript $(N)$ indicates the number of cells in the system;
the coefficients $a_1$ and $a_2$ are assumed to be constant.
For isolated systems, the conservation constraint $\sum_i^N \delta E_i =\sum_i^N \rho c_V\delta T_i = 0$, implies that $\sum_i^N \delta T_i = 0$. 
Summing Eq.~\eqref{eq:ansatz} then gives $a_2 = - a_1/N$.
Since the conservation correction vanishes in the limit $N \rightarrow \infty$; i.e.,
\begin{equation}
    \lim_{N\rightarrow\infty} \lrangleN{\delta T_i \delta T_j} 
    = \lrangle{\delta T^2} \delta_{ij} \, ,
\end{equation}
we have
\begin{equation}
    \lrangleN{\delta T_i^2} = \left(1 - \frac1N\right) \lrangle{\delta T^2} \, .
\label{eq:ConservationMom2}
\end{equation}
In other words, the variance of temperature fluctuations is reduced by a factor of $1/N$ due to conservation.
This analysis also gives $\lrangleN{\delta T_i \delta T_j} = -\lrangle{\delta T^2}/N$ for $i \neq j$, that is, fluctuations in different cells are correlated due to conservation.
This correlation due to conservation, often referred to as a finite-size correction, is well-known~(e.g.~\cite{Garcia_PRA_87}) and distinct from the long-ranged correlations in non-equilibrium systems.

Using similar arguments, we write
\begin{equation}
    \lrangleN{\delta T_i \delta T_j \delta T_k} 
    = a_1 \delta_{ij}\delta_{jk} + a_2(\delta_{ij} + \delta_{ik} + \delta_{jk}) + a_3 \, ,
\end{equation}
to find $a_2 = - a_1/N$ and $a_3 = -2 a_2/N$ so for $N > 2$,
\begin{equation}
    \lrangleN{\delta T_i^3} = \left(1 - \frac3N + \frac{2}{N^2}\right) \lrangle{\delta T^3} .
\label{eq:ConservationMom3}
\end{equation}
Additionally, we find that
\begin{equation}
    \lrangleN{\delta T_i^2 \delta T_j} 
    = -\frac{N-2}{N^2} \lrangle{\delta T^3} \, ,
    \quad
    \lrangleN{\delta T_i \delta T_j \delta T_k} 
    = \frac{2}{N^2} \lrangle{\delta T^3} \, ,    
\end{equation}
for $i \neq j \neq k$.

Finally, the analysis for the fourth central moment is somewhat complicated (see Appendix~\ref{app:4thMom}) with the final result being
\begin{subequations}
\begin{align}
    \lrangleN{\delta T_i^4} 
    &= \left(
        3 + \exkurt - \frac{6 + 4\exkurt}{N} + \frac{3 + 6\exkurt}{N^2} - \frac{3\exkurt}{N^3} 
    \right) \lrangle{\delta T^2}^2 \, ,
    \\
    \lrangleN{\delta T_i^3 \delta T_j} 
    &= \left(
        -\frac{3 + \exkurt}{N} +\frac{3 + 3\exkurt}{N^2} - \frac{3\exkurt}{N^3} 
    \right) \lrangle{\delta T^2}^2 \, ,
    \\
    \lrangleN{\delta T_i^2 \delta T_j^2} 
    &= \left(
        1 - \frac{2}{N} + \frac{3 + 2\exkurt}{N^2} - \frac{3\exkurt}{N^3} 
    \right) \lrangle{\delta T^2}^2 \, ,
    \\
    \lrangleN{\delta T_i^2 \delta T_j \delta T_k} 
    &= \left(
        -\frac{1}{N} + \frac{3 +\exkurt}{N^2} - \frac{3\exkurt}{N^3} 
    \right) \lrangle{\delta T^2}^2 \, ,
    \\
    \lrangleN{\delta T_i \delta T_j \delta T_k \delta T_l} 
    &= \left(
        \frac{3}{N^2} - \frac{3\exkurt}{N^3} 
    \right) \lrangle{\delta T^2}^2 \, ,
\end{align}
\label{eq:ConservationMom4} 
\end{subequations}
where $\exkurt = 12/{f\mathcal{N}}$ is the excess kurtosis for the gamma distribution.

It should be emphasized that these conservation corrections are physical effects. 
They simply reflect that for an isolated system represented in terms of a finite number of cells $\sum_i \delta T_i = 0$. 
Although this conservation correction is quite small for $N \gtrsim O(100)$, it can nevertheless be comparable to numerical modeling errors, as discussed in Section~\ref{sec:ModelingErrors}.

\section{Finite Volume Formulation}

To solve the stochastic heat equation (SHE) numerically, we introduce a finite-volume scheme that discretizes the equation in space and time. We break the domain into finite volume cells, $C_i = (i \Delta x, (i+1) \Delta x )$ and 
and let $t^n = n \Delta t$.
Our aim is to assess how well such numerical methods capture the central moments of the temperature distribution.
We let $T_i^n$ represent the value on $C_i$ at time $t^n$,   
for $i=0,\ldots,N-1$ where $N$ is the number of cells. 
For the temporal integration we take the simplest stochastic ODE scheme, namely Euler--Maruyama or Forward Euler~\cite{kloeden2012numerical}.
Although Euler--Maruyama is only first-order in time, we have focused on it here because it is relatively simple to analyze and it is consistent with It\^{o} stochastic calculus.
Higher-order schemes are briefly discussed in Section~\ref{sec:HigherOrderSchemes}.
Using this scheme 
\begin{equation}
\begin{split}
    T_i^{n+1} 
    &= T_i^n + \frac{\kappa \Delta t}{\Delta x^2} \left(T_{i+1}^n - 2 T_i^n  + T_{i-1}^n\right)
    + \frac{\noisecoeff \Delta t}{\Delta x} 
    \left(T_{i+\half}^n \tilde{Z}_{i+\half}^n - T_{i-\half}^n \tilde{Z}_{i-\half}^n\right) \\
    &= T_i^n + \frac{\Delta t}{\Delta x} \left( F_{i+\half}^n - F_{i-\half}^n \right) \, ,
\end{split}
\label{eq:fvdisc}    
\end{equation}
where the total flux, $F$, is the sum of the deterministic component, $\overline{F}_{i+\half}^n$, and a stochastic component, $\tilde{F}_{i+\half}^n$, given by 
\begin{equation}
    \overline{F}_{i+\half}^n = \frac{T_{i+1}^n-T_i^n}{\Delta x} \, ,
    \quad
    \tilde{F}_{i+\half}^n = \noisecoeff\, T_{i+\half}^n \tilde{Z}_{i+\half}^n \, ,
\end{equation}
respectively.
For the temperature on a face, we use the arithmetic average, $T_{i+\half}^n = (T_{i+1}^n + T_i^n)/2$. 
Numerically, the noise is generated as
\begin{equation}
    \tilde{Z}_{i+\half}^n = \frac{1}{\sqrt{V \Delta t}} \mathfrak{N}_{i+\half}^n \, ,
\label{eq:ZmathfrakN}
\end{equation}
where $V = \mathcal{A} \Delta x$ is the volume of a grid cell and $\mathfrak{N}$ are independent Gaussian pseudo-random numbers with zero mean and unit variance.

It is important to emphasize that our stochastic heat equation has \emph{multiplicative} noise. 
If the noise is made additive, say by setting $T_{i\pm\half}^n$ to the equilibrium temperature, then the correct equilibrium variance is obtained, but \emph{not} the higher order moments since the distribution of temperature for additive noise is Gaussian~\cite{Donev_CAMCOS_10}.

We consider three types of boundary conditions: periodic, Neumann, and Dirichlet.
The implementation of periodic boundary conditions is straightforward. 
For homogeneous Neumann boundary conditions, the flux of energy is zero at the boundary, so we set $F_{-\half}^n = F_{N-\half}^n = 0$ in the discretization Eq.~\eqref{eq:fvdisc} at an adiabatic boundary.
We will not consider inhomogeneous Neumann BCs (i.e., non-zero heat flux) since our attention is restricted to systems at thermodynamic equilibrium.

For Dirichlet boundary conditions, we specify $T$ on the faces of cells at the boundary, which is reflected in modifications to both the deterministic and stochastic fluxes at the boundary.
For the left and right boundary conditions, $T_\mathrm{L} = T(x=0)$ and $T_\mathrm{R} = T(x=L)$, the deterministic fluxes are written as
\begin{equation}    
    \overline{F}_{-\half}^n = \frac{T_0^n-T_\mathrm{L}}{\Delta x / 2} \, , 
    \quad
    \overline{F}_{N-\half}^n = \frac{T_\mathrm{R} - T_{N-1}^n}{\Delta x / 2} \, ,
\end{equation}
respectively.  
To ensure that the discrete system captures the correct variance at equilibrium, the stochastic fluxes at the left and right faces are modified to 
\begin{equation}    
    \tilde{F}_{-\half}^n 
    = \sqrt{2} \noisecoeff \mathcal{D}(T_\mathrm{L},T_0^n) \tilde{Z}_{-\frac12}^n \, ,
    \quad
    \tilde{F}_{N-\half}^n 
    = \sqrt{2} \noisecoeff \mathcal{D}(T_\mathrm{R},T_{N-1}^n) \tilde{Z}_{N-\frac12}^n \, ,    
\end{equation}
respectively.
For the specification of the stochastic flux with Dirichlet boundary conditions, we consider $\mathcal{D}$ of the form
\begin{equation}
    \mathcal{D}_\zeta(T_\mathrm{L},T) = \zeta T + (1 - \zeta) T_\mathrm{L} \, ,
\label{eq:Dzeta}
\end{equation}
with either $\zeta = 0$ or $\zeta = \frac12$.
\footnote{Note that the case $\zeta = 1$ was also investigated and found to give inferior results with regard to the correctness of the moments.}
Earlier work~\cite{Garcia_JSP_87,garcia2025introCFHD} used $\mathcal{D}_0$ since it gives accurate results for the variance. However, as shown in Section~\ref{sec:ModelingErrors}, using $\mathcal{D}_\frac12$ gives better results for the higher moments.

\section{Numerical Tests}\label{sec:NumericalTests}

This section discusses results from numerical tests in which $\lrangle{\delta T^m}$ was measured in simulations of the stochastic heat equation using the finite volume algorithm described in the previous section. 
These simulation measurements are compared with the central moments given in Section~\ref{sec:MomentsTheory}.
Unless stated otherwise, the simulation parameters are: $\overline{T} = 300~\mathrm{K}$; $f = 6$; $\Delta x = 1\times10^{-9}~\mathrm{m}$; $\mathcal{A} = 2.5\times10^{-19}~\mathrm{m}^2$, $\lambda = 70~\mathrm{\frac{W}{m\cdot K}}$.

Here we consider mesoscopic systems, by which we mean that the number of molecules per cell is small enough to have noticeable fluctuations ($\mathcal{N} \leq 100$) but not so small that the fluctuating hydrodynamic formulation is inappropriate ($\mathcal{N} \geq 10$).
In this section, we also consider systems with large numbers of grid cells ($N \gg 1$), specifically, we take either $N = 100$ or $N = 1000$.
For these large $N$, the conservation correction is relatively small. 
We view the discrepancies between the theoretical moments and the simulations as primarily modeling errors, typically $O(1/f\mathcal{N})$ for the central moments.

\begin{figure}[htbp]
    \centering
    \includegraphics[width= \textwidth]{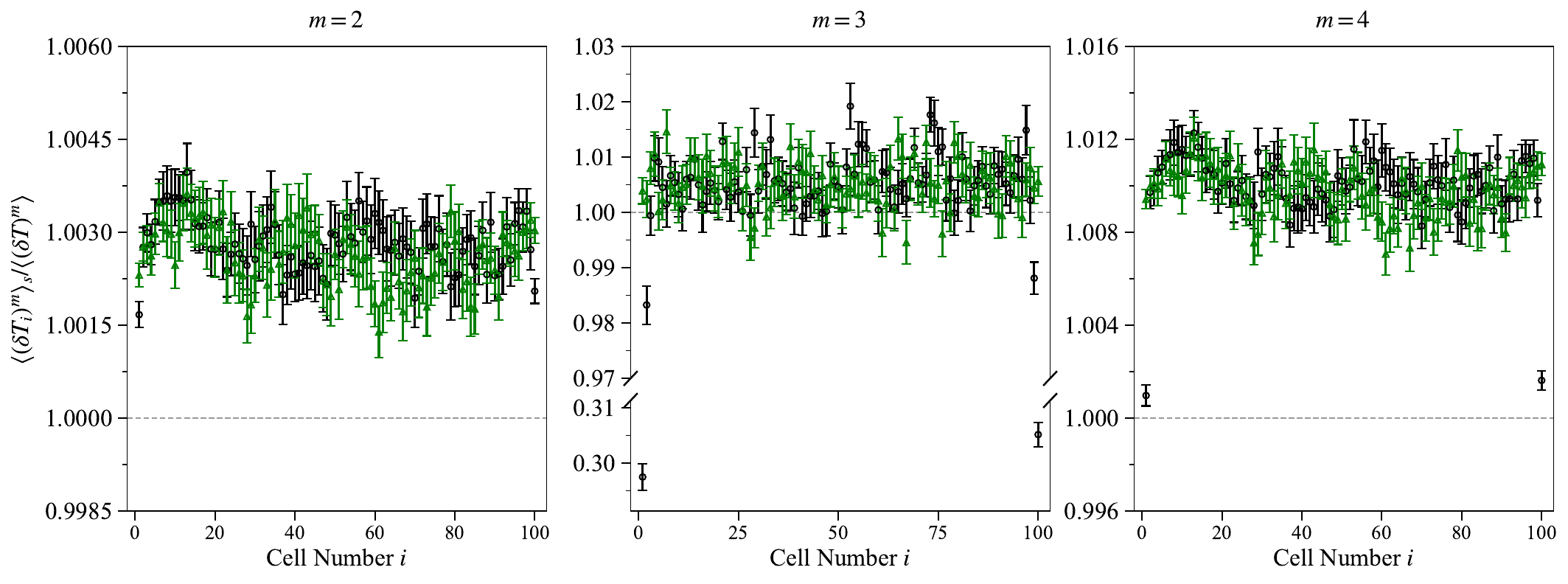}
    \caption{Measured normalized central moments $\lrangleS{ \delta T^m_i } / \lrangle{ \delta T^m_i}$ versus cell index $i$ for $m=2$ (left), $m=3$ (center), and $m=4$ (right). Systems have $N = 100$ cells and Dirichlet boundary conditions with either $\zeta = 0$ (black circles) or $\zeta = \frac12$ (green triangles).}
    \label{fig:MomentsDirichlet}
\end{figure}

Figure~\ref{fig:MomentsDirichlet} compares the measured normalized central moments, $\lrangleS{ \delta T^m_i } / \lrangle{ \delta T^m_i}$, for the two options for Dirichlet boundary conditions with $T_\mathrm{L} = T_\mathrm{R} = \overline{T}$.
For these runs, the time step was $\Delta t = 1.86 \times 10^{-16}$ s, $N = 100$, and $\mathcal{N} = 100$.
The traditional choice, $\zeta = 0$, performs well for the second and fourth moments but has a very large error in the third moment for cells next to the boundary.
By contrast, the other option, $\zeta = \frac12$, gives good results (fractional difference under one percent) for all three moments.
In particular, with $\zeta = \frac12$ the error in the second moment is roughly 0.3\% and the error in the third and fourth moments are $\approx$ 1\%.

\begin{figure}[htbp]
    \centering
    \includegraphics[width= \textwidth]{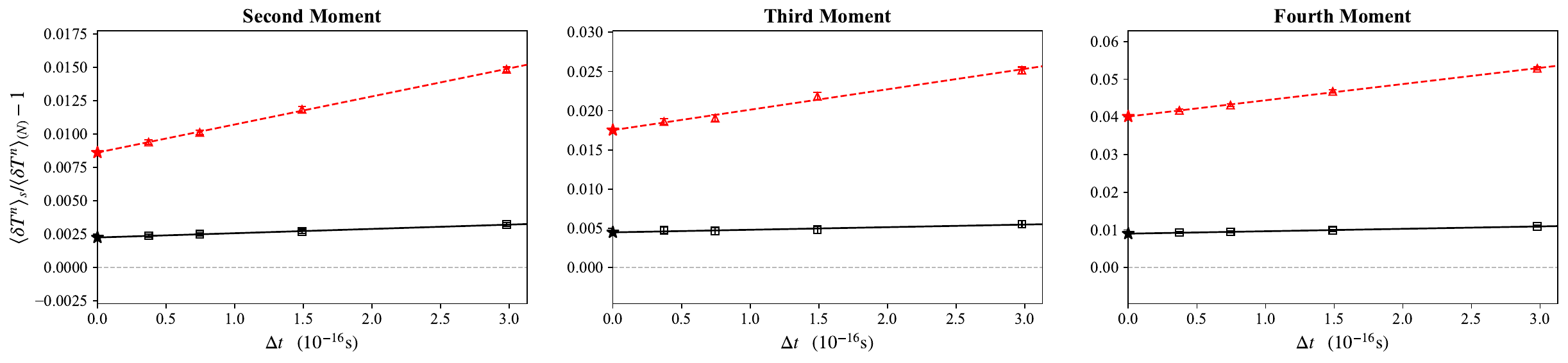}
    \caption{Measured fractional difference, $\lrangleS{ \delta T^m } / \lrangle{ \delta T^m } - 1$, for the second (left), third (center), and fourth (right) central moments versus time step $\Delta t$ for Dirichlet boundary conditions. Symbols are from individual simulations and lines are fits for $N = 100$ and either $\mathcal{N} = 20$ (red) or $100$ (black).}
    \label{fig:DirichletTimestep}
\end{figure}

We now look more closely at how these simulation results for Dirichlet boundary conditions depend on  $\Delta t$ and $\mathcal{N}$.
For Euler--Maruyama, the truncation error in the temperature variance is $O(\Delta t / \Delta x^2)$~\cite{Donev_CAMCOS_10} so we performed a number of simulations at several different time steps and extrapolated those results to the limit of vanishing $\Delta t$.
Figure~\ref{fig:DirichletTimestep} shows the measured fractional difference, $\lrangleS{ \delta T^m } / \lrangle{ \delta T^m } - 1$, versus $\Delta t$ for $m= 2, 3, 4$; here $\lrangleS{ \delta T^m }$ indicates an average taken over an ensemble of simulations and the $\lrangle{ \delta T^m }$'s are given by Eq.~\eqref{eq:T_moments_fN}. 
From these results we see that the fractional differences observed in Fig.~\ref{fig:MomentsDirichlet} are not due to truncation error for finite time step.
We also notice that this fractional difference in the central moments decreases with increasing $\mathcal{N}$; this will be analyzed in Section~\ref{sec:ModelErrorsDirichlet}.

\begin{figure}[htbp]
    \centering
    \includegraphics[width= 1\textwidth]{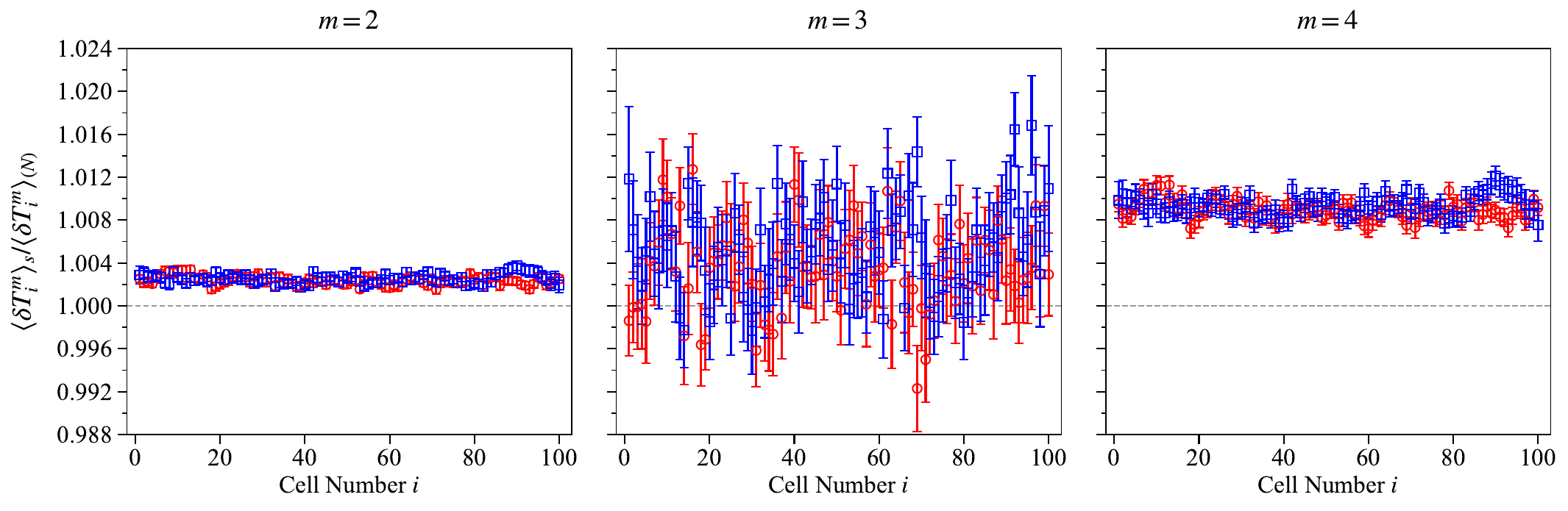}
    \caption{Measured normalized central moments $\lrangleS{ \delta T^m_i } / \lrangleN{ \delta T^m_i}$ versus cell index $i$ for $m=2$ (left), $m=3$ (center), and $m=4$ (right). Systems have $N = 100$ cells, $\mathcal{N} = 100$, and either periodic (red circles) or Neumann (blue squares) boundary conditions.
    Dotted line is $\lrangleS{ \delta T^m_i } / \lrangleN{ \delta T^m_i} = 1$.}
    \label{fig:MomentsBCs}
\end{figure}

Figure~\ref{fig:MomentsBCs} shows the measured normalized central moments, $\lrangleS{ \delta T^m_i } / \lrangleN{ \delta T^m_i}$,  for isolated systems, specifically systems with either periodic or homogeneous Neumann boundary conditions. 
Here $\lrangleS{ \delta T^m }$ indicates an average taken over an ensemble of simulations and $\lrangleN{ \delta T^m }$ are given by Eqs.~\eqref{eq:ConservationMom2}, \eqref{eq:ConservationMom3}, \eqref{eq:ConservationMom4}.
For these runs, the time step was $\Delta t = 1.86 \times 10^{-16}$ s, $N = 100$, and $\mathcal{N} = 100$.
The general agreement with the theory is good (fractional difference below 1\%) for both boundary conditions, and the cells near the Neumann boundaries are indistinguishable from cells in the interior.

\begin{figure}[htbp]
    \centering
    \includegraphics[width= \textwidth]{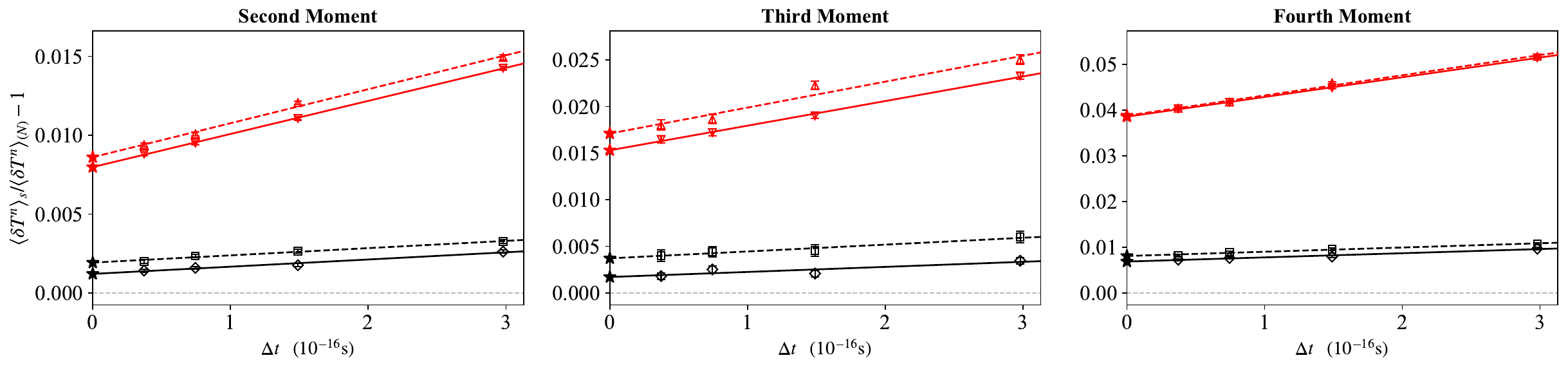}
    \caption{Measured fractional difference, $\lrangleS{ \delta T^m } / \lrangleN{ \delta T^m } - 1$, for the second (left), third (center), and fourth (right) central moments versus time step $\Delta t$ for periodic boundary conditions. Symbols are from individual simulations and lines are fits for $\mathcal{N} = 20$ (red) or $100$ (black) and $N = 100$ (dashed lines) or $1000$ (solid lines).}
    \label{fig:PeriodicTimestep}
\end{figure}

Figure~\ref{fig:PeriodicTimestep} shows how the simulation results for periodic boundary conditions depend on $N$, $\mathcal{N}$, and $\Delta t$; similar results are found for Neumann boundary conditions.
We see that there is a small error in the moments that decreases linearly with decreasing time step, as expected, however the fractional difference does not go to zero as $\Delta t \rightarrow 0$.
This fractional difference does decrease with increasing $\mathcal{N}$ and with increasing $N$; 
these constitute modeling errors that are discussed in Section~\ref{sec:ModelErrorsPeriodic}.

\begin{figure}[htbp]
    \centering
    \includegraphics[width=\linewidth]{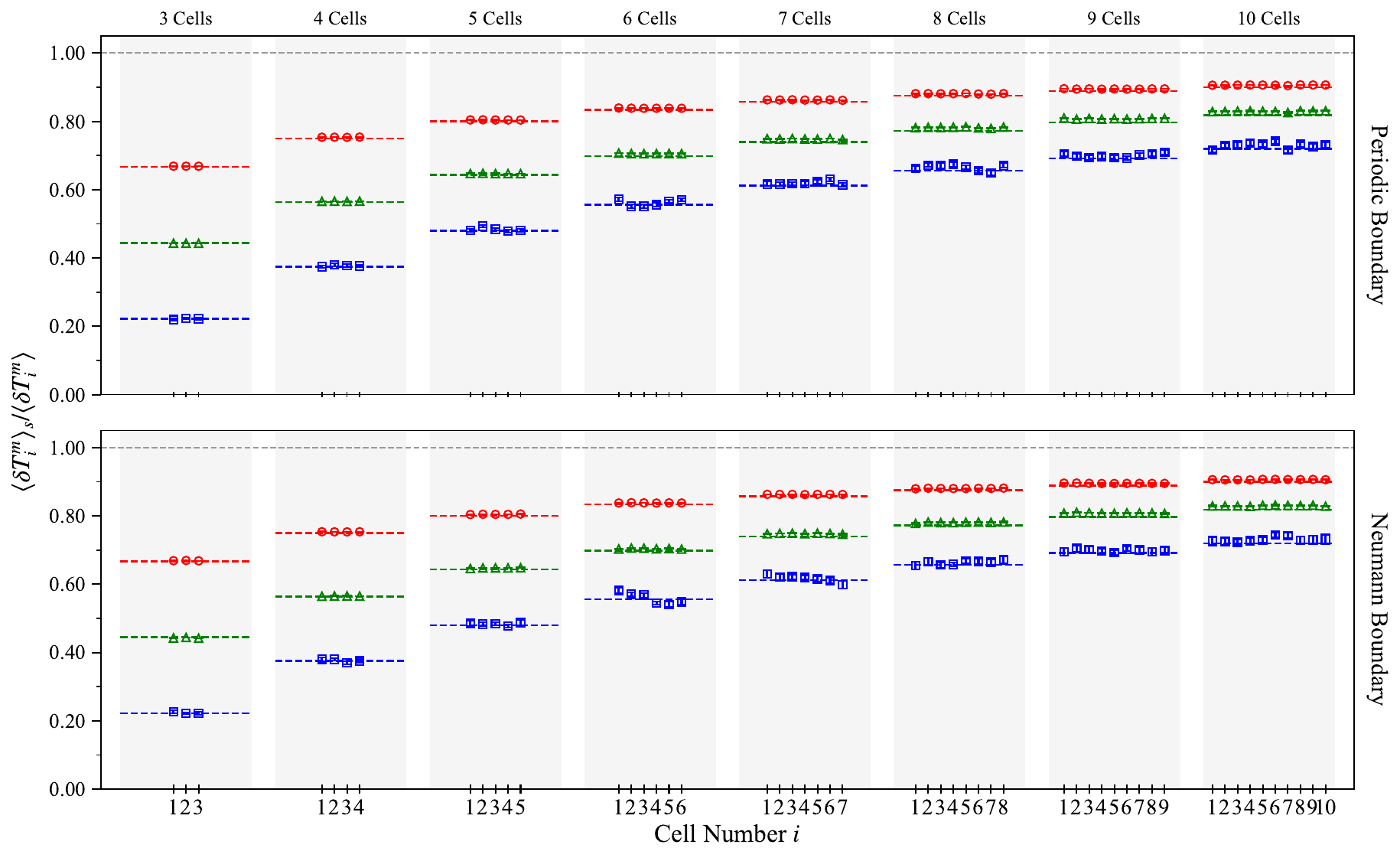}
    \caption{Normalized measured moments, $\lrangleS{\delta T_i^m}/\lrangle{\delta T_i^m}$ for $m = 2$ (red circles), 3 (blue squares), and 4 (green triangles). Measurements are from individual cells in simulations with $N=3, \ldots, 10$ cells and either periodic (top panel) or Neumann (bottom panel) boundary conditions. The red, blue, and green dashed lines are given by Eqs.~\eqref{eq:ConservationMom2}, \eqref{eq:ConservationMom3}, and \eqref{eq:ConservationMom4}, respectively.}
    \label{fig:conservation_correction}
\end{figure}

To provide a more detailed characterization of the conservation correction, we examine its impact for systems with small numbers of cells.
Specifically, Figure~\ref{fig:conservation_correction} shows simulation results for systems with $N = 3, \ldots, 10$ cells for both periodic and Neumann boundary conditions.
For these runs, the time step was $\Delta t = 9.88 \times 10^{-18}$ s, and $\mathcal{N} = 100$; statistics were accumulated over $4 \times10^{9}$ time steps.
These results verify that even for systems with a small number of cells the simulation measurements are in good agreement with the expressions for moments with conservation corrections, $\lrangleN{ \delta T^m}$, given Eqs.~\eqref{eq:ConservationMom2}, \eqref{eq:ConservationMom3}, and \eqref{eq:ConservationMom4}.
It should be noted that the agreement between $\lrangleS{ \delta T^m}$ and $\lrangleN{ \delta T^m}$ is not as good when $\mathcal{N}\lessapprox O(10)$; this was already seen in Fig.~\ref{fig:PeriodicTimestep} and it is analyzed in Section~\ref{sec:ModelErrorsPeriodic}.

\section{Modeling Errors}\label{sec:ModelingErrors}

From Figures~\ref{fig:DirichletTimestep} and \ref{fig:PeriodicTimestep} we see that the fractional difference between the central moments, as measured in the simulations versus those predicted by theory, depends on both $\mathcal{N}$ and $N$, even in the limit of vanishing time step.
In this section we show that modeling the stochastic heat equation as a system of stochastic ordinary differential equations with multiplicative noise and discretizing those equations in space using standard finite volume schemes results in central moments that differ from Eq.~\eqref{eq:T_moments_fN} by terms of $O(1 / (f\mathcal{N}))$. 
Separate from the conservation correction, for isolated systems, this modeling error also depends weakly on $N$, the number of cells.

\subsection{Dirichlet Boundary Conditions}\label{sec:ModelErrorsDirichlet}

\subsubsection{Single cell case}\label{sec:ModelErrorsDirichletSingle}

We start with a simple scenario for which the higher moments with Dirichlet boundary conditions are easily calculated analytically and expressed compactly.
Consider a system with one cell with the left face being a thermal wall (Dirichlet boundary at temperature $T_\mathrm{L}$) and the right face being an adiabatic wall (Neumann boundary with zero heat flux). 
The Euler--Maruyama scheme for this system is
\begin{equation}
    T_0^{n+1} 
    = T_0^n 
    - \frac{2\kappa\Delta t}{\Delta x^2} \left(T_{0}^n - T_\mathrm{L}\right)
    - \frac{\sqrt{2}\noisecoeff\Delta t}{\Delta x} \mathcal{D}(T_\mathrm{L}, T_0^n) \tilde{Z}_{-\half}^n \, .
\label{eq:onecellorig}
\end{equation}
If we define $T' = T_0^{n+1}$, $T = T_0^n$, $Z = \tilde{Z}_{-\frac12}$ we can write Eq.~\eqref{eq:onecellorig} as
\begin{equation}
    T' = T + A(2 T_\mathrm{L} - 2T) -\sqrt{2} B\, \mathcal{D}_\zeta(T_\mathrm{L}, T)\, Z \, ,
\label{eq:OneCellFE}
\end{equation}
where  $A = \kappa \Delta t/\Delta x^2$, and $B = \noisecoeff \Delta t / \Delta x$.
Note that $\lrangle{ Z^2 } = 1/(\Delta t V)$ where $V = \mathcal{A} \Delta x$ is the volume of the single-cell system.
If we define $\delta T = T -\lrangle{T}$, Eq.~\eqref{eq:OneCellFE} becomes
\begin{equation}
    \delta T' = \delta T - 2A\delta T - \sqrt{2} B\, \mathcal{D}_\zeta(T_\mathrm{L}, T)\, Z  \, ,
\label{eq:SHE_FE_OneCell}
\end{equation}
which allows us to obtain central moments by solving
\begin{equation}
    \lrangleDT{(\delta T)^m} 
    = \lrangleDT{(\delta T - 2A \delta T  - \sqrt{2} B\, \mathcal{D}_\zeta(T_\mathrm{L}, T)\, Z )^m } \, ,
\end{equation}
since $\lrangleDT{(\delta T')^m}  = \lrangleDT{\delta T^m}$.
Here the subscript ``$\Delta t$'' is to indicate that the average is being evaluated at a finite time step from the model equations, that is, the discretized form of the stochastic heat equation.
We take $T_\mathrm{L} = \overline{T}$ so that, since $\lrangleDT{Z} = 0$, we have $\lrangleDT{ \delta T} = 0$ for all $\Delta t$ and $\zeta$. 

Here we do not want to consider the $O(\Delta t/ \Delta x^2)$ temporal truncation error terms arising from the Euler--Maruyama discretization. 
Hence, we define
\begin{equation}
    \lrangleM{(\delta T)^m} = \lim_{\Delta t\rightarrow 0} \lrangleDT{\delta T^m} \, ,
\end{equation}
where the subscript ``m'' indicates that the average is evaluated for the model equations in the limit of vanishing time step.
For $\zeta = 0$, the temperature variance is then
\begin{equation}
\begin{split}
    \lrangleM{\delta T^2} 
    &=\lim_{\Delta t\rightarrow 0} \lrangleDT{\delta T^2}
    = \lim_{\Delta t\rightarrow 0} \frac{B^2}{2A(1-A) V\Delta t} \Tbar^2
    \\
    &= \lim_{\Delta t\rightarrow 0} \left(
        \frac{2 \Tbar^2}{f\mathcal{N}} + O(\Delta t/\Delta x^2) 
    \right )
    = \frac{2\Tbar^2}{f\mathcal{N}} \, , 
\end{split}
\end{equation}
which is exactly the theoretical result from statistical mechanics. 
However, for $\zeta = 0$ we find, after dropping the truncation error terms, that $\lrangleM{ \delta T^3 } = 0$, which does \emph{not} agree with the theoretical result since it gives $\mathrm{Skew}[T] = 0$ while we see from Eq.~\eqref{eq:T_SkewKurt} that the skewness should be non-zero.
In fact, Eq.~\eqref{eq:SHE_FE_OneCell} using $\mathcal{D}_0$, leads to a linear stochastic ODE with \emph{additive} noise.
Thus, we expect the resulting distribution for $\delta T$ to be Gaussian and, as such, the skewness is zero and the kurtosis is 3.

By comparison, for $\zeta = \frac12$, again neglecting the Euler--Maruyama truncation error, the variance is
\begin{equation}
\begin{split}
    \lrangleM{\delta T^2} 
    &= 
    \frac{2\Tbar^2}{f\mathcal{N}(1 - \frac{1}{2f\mathcal{N}})} =
    \lrangle{\delta T^2}\frac{1}{(1 - \frac{1}{2f\mathcal{N}})} 
    \\ 
    &\approx \lrangle{\delta T^2} \left(
        1 + \frac{1}{2f\mathcal{N}} + O\left(\frac{1}{(f\mathcal{N})^2}
    \right)\right) \, ,
\end{split}
\label{eq:OneCellMom2EM}
\end{equation}
so the difference from Eq.~\eqref{eq:T_moments_fN} is $O(1/(f\mathcal{N}))$.
The third central moment is
\begin{equation}
    \lrangleM{\delta T^3} = 
    \frac{8\Tbar^3}{(f\mathcal{N})^2 \left(1 - \frac{3}{2f\mathcal{N}} - \frac{1}{2(f\mathcal{N})^2}\right)}
    \approx 
    \lrangle{\delta T^3} \left(
        1 + \frac{3}{2f\mathcal{N}}
    \right) \, ,
\label{eq:OneCellMom3EM}
\end{equation}
which agrees with Eq.~\eqref{eq:T_moments_fN} up to a correction factor of $O(1/(f\mathcal{N}))$.
The fourth central moment is 
\begin{equation}
\begin{split}
    \lrangleM{ \delta T^4 } 
    &= \frac{(12+\frac{36}{f\mathcal{N}})\Tbar^4}
        {(f\mathcal{N})^2(1-\frac{3}{f\mathcal{N}}+\frac{11}{4 (f\mathcal{N})^2}-\frac{3}{4 (f\mathcal{N})^3})} \\
    &= \lrangle{\delta T^4} \frac{1+\frac{3}{f\mathcal{N}}}
    {(1+\frac{4}{f\mathcal{N}})(1-\frac{3}{f\mathcal{N}}+\frac{11}{4 (f\mathcal{N})^2}-\frac{3}{4 (f\mathcal{N})^3})} \\
    &\approx \lrangle{\delta T^4} \left(
        1 + \frac{2}{f\mathcal{N}}
    \right) \, ,
\end{split}
\label{eq:OneCellMom4EM}
\end{equation}

In summary, for the simple case of a single cell, the Dirichlet boundary function $\mathcal{D}_\frac12 = \frac12 (T_\mathrm{L} + T)$ gives the accurate results (with an $O(1/(f\mathcal{N}))$ modeling error) for the second through fourth central moments. 
The traditional alternative, $\mathcal{D}_0 = T_\mathrm{L}$, while giving the correct second moment does very poorly for the third moment.

\subsubsection{Multi-Cell Cases}\label{sec:ModelErrorsDirichletMulti}

The above analysis can be extended to domains with 2, 3, or 4 cells with a Dirichlet boundary on the left and a Neumann boundary on the right. 
In these cases, the results differ from cell to cell. 
Rather than treating each case separately, we employ the general multi-cell multiplicative-noise formulation described in Appendix~\ref{app:multiplicative_noise}, from which the moment equations can be derived systematically.

For two cells, we obtain the following second moments:
\begin{subequations}
\begin{align}
    \lrangleM{\delta T_0^2} 
    &= \lrangle{\delta T^2} \frac{1}{1-\frac{5}{8}\frac{1}{f\mathcal{N}}} 
    \approx \lrangle{\delta T^2}\left(1+\frac{5}{8}\frac{1}{f\mathcal{N}}\right) \, , 
    \\
    \lrangleM{\delta T_1^2} 
    &= \lrangle{\delta T^2} \frac{1+\frac{1}{4}\frac{1}{f\mathcal{N}}}{1-\frac{5}{8}\frac{1}{ f\mathcal{N}}} 
    \approx \lrangle{\delta T^2}\left(1+\frac{7}{8}\frac{1}{f\mathcal{N}} \right) \, , 
\end{align}
\end{subequations}
where $i = 0$ is the cell adjacent to the Dirichlet boundary and $i = 1$ is the cell adjacent to the Neumann boundary.
There is also a cross correlation between these two cells
\begin{equation}
    \lrangleM{\delta T_0 \delta T_1} 
    = \lrangle{\delta T^2} \frac{1}{5-8f\mathcal{N}} 
    \approx \lrangle{\delta T^2}\left(
        -\frac{1}{8f\mathcal{N}} 
    \right) \, . 
\end{equation}
For higher-order moments, the expression becomes much more complex, so we consider only the leading order terms in $1/(f\mathcal{N})$.
For the third-order moments this gives
\begin{equation}
    \lrangleM{\delta T_0^3} 
    \approx \lrangle{\delta T^3} \left(1 + \frac{215}{136} \frac{1}{f\mathcal{N}}\right) \, ,
    \quad
    \lrangleM{\delta T_1^3} 
    \approx \lrangle{\delta T^3} \left(1 + \frac{229}{136} \frac{1}{f\mathcal{N}}\right) \, ,
\end{equation}
with cross correlation terms
\begin{equation}
    \lrangleM{\delta T_0^2 \delta T_1} 
    \approx \lrangle{\delta T^3} \left( 
        -\frac{1}{136} \frac{1}{f\mathcal{N}} 
    \right ) \, ,
    \quad
    \lrangleM{\delta T_0\delta T_1^2} 
    \approx \lrangle{\delta T^3} \left( 
        -\frac{9}{136} \frac{1}{f\mathcal{N}} 
    \right) \, .
\end{equation}
Finally, for fourth moments, we obtain
\begin{equation}
    \lrangleM{\delta T_0^4} 
    \approx \lrangle{\delta T^4} \left(1 + \frac{489}{224} \frac{1}{f\mathcal{N}}\right) \, ,
    \quad
    \lrangleM{\delta T_1^4} 
    \approx \lrangle{\delta T^4} \left(1 + \frac{529}{224} \frac{1}{f\mathcal{N}}\right) \, ,
\end{equation}
with cross correlations 
\begin{equation}
\begin{split}
    \lrangleM{\delta T_0^3 \delta T_1} 
    &\approx \lrangle{\delta T^4} \left( 
        -\frac{73}{224} \frac{1}{f\mathcal{N}} 
    \right) \, , 
    \quad
    \lrangleM{\delta T_0\delta T_1^3} 
    \approx \lrangle{\delta T^4} \left( 
    - \frac{115}{224} \frac{1}{f\mathcal{N}} 
    \right) \, ,
    \\
    \lrangleM{\delta T_0^2 \delta T_1^2} 
    &\approx \frac{\lrangle{\delta T^4}}{3} \left( 1 - \frac{323}{224} \frac{1}{f\mathcal{N}}\right) \, .
\end{split}
\end{equation}

Analogously, for three cells we obtain the following second moments:
\begin{subequations}
\begin{align}
    \lrangleM{\delta T_0^2} 
    &\approx \lrangle{\delta T^2}\left(1+\frac{33}{52}\frac{1}{f\mathcal{N}} \right) \, ,
    \\
    \lrangleM{\delta T_1^2} 
    &\approx \lrangle{\delta T^2}\left(1+\frac{12}{13}\frac{1}{f\mathcal{N}} \right) \, ,
    \\
    \lrangleM{\delta T_2^2} 
    &\approx \lrangle{\delta T^2}\left(1+\frac{49}{52}\frac{1}{f\mathcal{N}} \right) \, ,
\end{align}
\end{subequations}
where $i = 0$ is the cell adjacent to the Dirichlet boundary, $i = 1$ is the center cell, and $i = 2$ is the cell adjacent to the Neumann boundary. 
In addition, we obtain
\begin{subequations}
\begin{align}
    \lrangleM{\delta T_0^3} 
    &\approx \lrangle{\delta T^3} \left(1 + \frac{4127}{2574} \frac{1}{f\mathcal{N}} \right) \, , 
    \\
    \lrangleM{\delta T_1^3} 
    &\approx \lrangle{\delta T^3} \left(1 + \frac{4619}{2574} \frac{1}{f\mathcal{N}} \right) \, ,
    \\
    \lrangleM{\delta T_2^3} 
    &\approx \lrangle{\delta T^3} \left(1 + \frac{4753}{2574} \frac{1}{f\mathcal{N}} \right) \, ,
\end{align}
\end{subequations}
for the third moments and
\begin{subequations}
\begin{align}
    \lrangleM{\delta T_0^4} 
    &\approx \lrangle{\delta T^4} \left(1 + \frac{6991}{3172} \frac{1}{f\mathcal{N}} \right) \, , 
    \\
    \lrangleM{\delta T_1^4} 
    &\approx \lrangle{\delta T^4} \left(1 + \frac{2020}{793} \frac{1}{f\mathcal{N}} \right) \, ,
    \\
    \lrangleM{\delta T_2^4} 
    &\approx \lrangle{\delta T^4} \left(1 + \frac{7743}{3172} \frac{1}{f\mathcal{N}} \right) \, .
\end{align}
\end{subequations}
for the fourth moments.
As in the two-cell case, cross correlations are correct to order $1/(f\mathcal{N})$.

Finally, for four-cell systems we have the second moments
\begin{subequations}
\begin{align}
    \lrangleM{\delta T_0^2} 
    &\approx \lrangle{\delta T^2} \left(1 + \frac{173}{272} \frac{1}{f\mathcal{N}} \right) \, ,
    \\
    \lrangleM{\delta T_1^2} 
    &\approx \lrangle{\delta T^2} \left(1 + \frac{127}{136} \frac{1}{f\mathcal{N}} \right) \, , 
    \\
    \lrangleM{\delta T_2^2} 
    &\approx \lrangle{\delta T^2} \left(1 + \frac{131}{136} \frac{1}{f\mathcal{N}} \right) \, ,
    \\
    \lrangleM{\delta T_3^2} 
    &\approx \lrangle{\delta T^2} \left(1 + \frac{263}{272} \frac{1}{f\mathcal{N}} \right) \, ,
\end{align}
\end{subequations}
and third moments
\begin{subequations}
\begin{align}
    \lrangleM{\delta T_0^3} 
    &\approx \lrangle{\delta T^3} \left(1 + \frac{1445193287}{899132176 } \frac{1}{f\mathcal{N}} \right) \, ,
    \\
    \lrangleM{\delta T_1^3} 
    &\approx \lrangle{\delta T^3} \left(1 + \frac{819624061}{449566088} \frac{1}{f\mathcal{N}} \right) \, , 
    \\
    \lrangleM{\delta T_2^3} 
    &\approx \lrangle{\delta T^3} \left(1 + \frac{851953641}{449566088} \frac{1}{f\mathcal{N}} \right) \, ,
    \\
    \lrangleM{\delta T_3^3} 
    &\approx \lrangle{\delta T^3} \left( 1 + \frac{1712762021}{899132176} \frac{1}{f\mathcal{N}} \right) \, ,
\end{align}
\end{subequations}
and fourth moments
\begin{subequations}
\begin{align}
    \lrangleM{\delta T_0^4} 
    &\approx \lrangle{\delta T^4} \left(1 + \frac{2288935876707}{1037199712768 } \frac{1}{f\mathcal{N}} \right) \, ,
    \\
    \lrangleM{\delta T_1^4} 
    &\approx \lrangle{\delta T^4} \left(1 + \frac{45437271084873}{17632395117056} \frac{1}{f\mathcal{N}} \right) \, ,
    \\
    \lrangleM{\delta T_2^4} 
    &\approx \lrangle{\delta T^4} \left(1 + \frac{46026326584681}{17632395117056} \frac{1}{f\mathcal{N}} \right) \, ,
    \\
    \lrangleM{\delta T_3^4} 
    &\approx \lrangle{\delta T^4} \left(1 + \frac{43785272852931}{17632395117056} \frac{1}{f\mathcal{N}} \right) \, .
\end{align}
\end{subequations}

\begin{figure}[htbp]
    \centering
    \subfloat[Second Moment\label{fig:second_mom_dirichlet}]
    { 
        \includegraphics[width=0.48\textwidth]{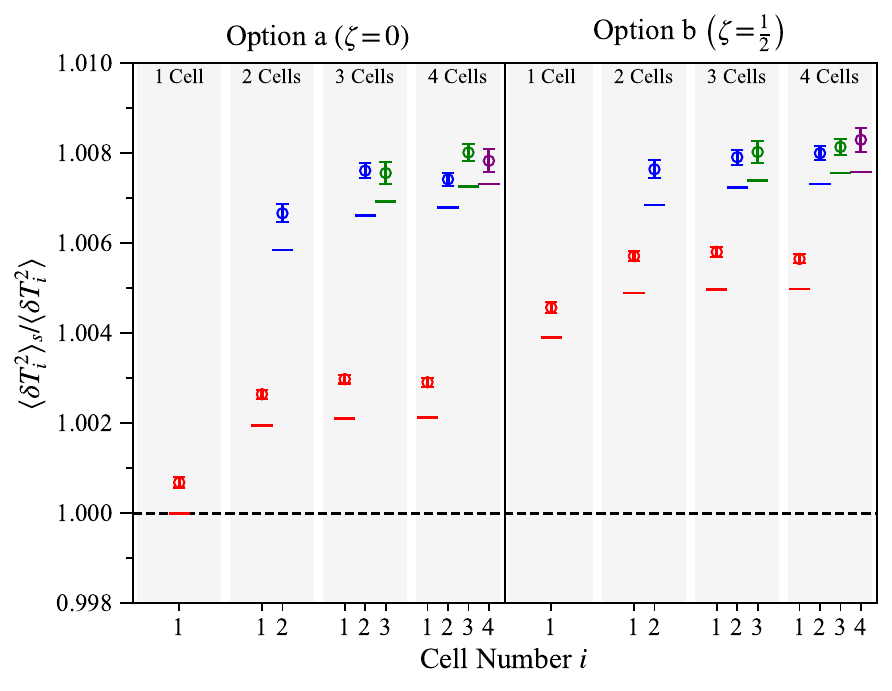} 
    }
    \vspace{0.3cm} 
    \subfloat[Third Moment\label{fig:third_mom_dirichlet}]{ 
        \includegraphics[width=0.48\textwidth]{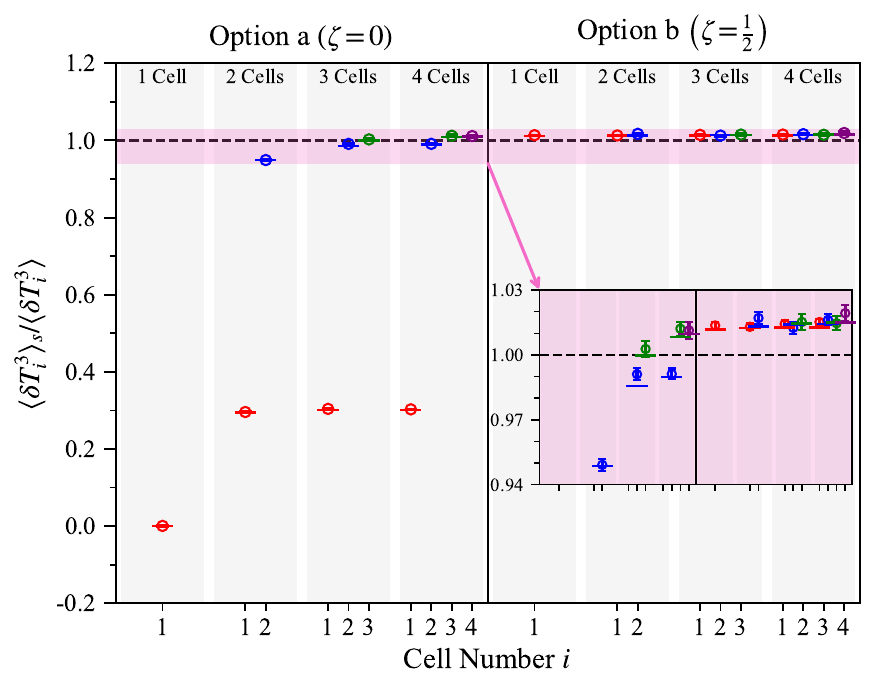} 
    } 
    \hfill 
    \subfloat[Fourth Moment\label{fig:fourth_mom_dirichlet}]{ 
        \includegraphics[width=0.48\textwidth]{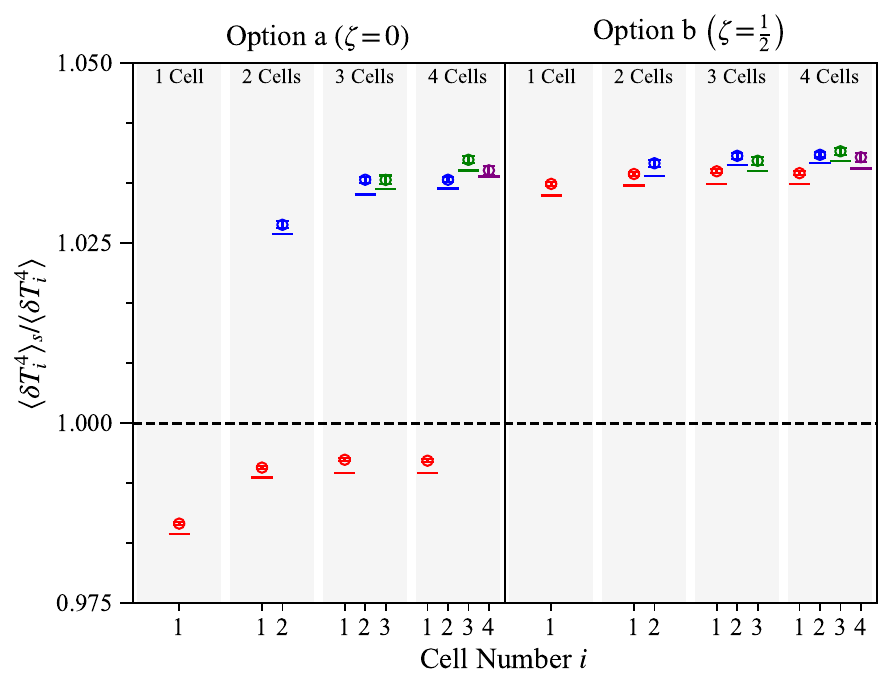} 
    }
    \caption{Comparison of central moments for $N=1,\ldots, 4$ cell systems with a Dirichlet boundary on the left and Neumann boundary on the right. Simulation results, $\lrangleS{ \delta T_i^m }/\lrangle{ \delta T_i^m }$, (markers) and the analytic results predicted using the model corrections, $\lrangleM{ \delta T_i^m }/\lrangle{ \delta T_i^m }$, (horizontal lines) are shown for the two Dirichlet boundary function options ($\zeta = 0$ and $\zeta = \frac12$). Panels show (a) the second moment, (b) the third moment and (c) the fourth moment.}
    \label{fig:dirichlet_moments}
\end{figure}

Figure~\ref{fig:dirichlet_moments} presents results from simulations that illustrate the model error for the two Dirichlet boundary options ($\zeta = 0$ and $\zeta = \frac12$).
As in the analysis above, the simulations imposed a Dirichlet boundary condition on the left side and a homogeneous Neumann (i.e., zero flux) condition on the right side.
To highlight its contribution to the model error, we take the number of particles per cell as $\mathcal{N} = 20$.
The figure shows that, in all cases, the theoretical predictions for the modeling error match the simulation results to within approximately 1\%.
The figure also shows that in most cases, even for $\mathcal{N}=20$, the simulations are in reasonable agreement with the theoretical values given by Eq.~\eqref{eq:T_moments_fN}.
The exception is the third moment in the cell next to the Dirichlet boundary for which the model error is greater than 50\% when $\zeta = 0$.

\subsection{Periodic Boundary Conditions}\label{sec:ModelErrorsPeriodic}

We now turn to the formulation of the model error results for periodic boundary conditions.
Using the framework described in Appendix~\ref{app:multiplicative_noise}, closed-form expressions can be obtained for systems with a small number of cells. 
In contrast to the Dirichlet case, translational symmetry implies that all cells are statistically equivalent, leading to substantially simpler expressions for the moments and correlations. 
Analytical results obtained for small values of $N$ reveal a clear dependence on both $N$ and $f\mathcal{N}$. 
Based on these results, we inferred the general forms for the second- and third-order moments presented below and subsequently verified them through numerical simulations of larger systems. 
We note that we were unable to obtain closed-form expressions for the fourth-order moments.

For the second moment we find that 
\begin{align}\label{eq:PeriodicModel2}
    \lrangleM{\delta T_i^2} 
    &= 
    \left(1-\frac{1}{N}\right)\frac{2 \Tbar^2} {f\mathcal{N}\left(1 - \frac{N-2}{N}\frac{1}{f\mathcal{N}}\right)}
    \approx \lrangleN{\delta T^2_i} \left(1 + \frac{N-2}{N }\frac{1}{f\mathcal{N}} 
    \right) \, ,
    \\
    \lrangleM{\delta T_i \delta T_j} 
    &= 
    -\frac{1}{N}\frac{ 2\Tbar^2} {f\mathcal{N} \left(1 - \frac{N-2}{N }\frac{1}{f\mathcal{N}}\right)} 
    \approx \lrangleN{\delta T_i \delta T_j} \left(1 + \frac{N-2}{N }\frac{1}{f\mathcal{N}} 
    \right) \, .
\end{align}
For the third moment we have
\begin{align}\label{eq:PeriodicModel3}
    \lrangleM {\delta T_i^3} 
    &= 
    \left( 1 - \frac{3}{N} + \frac{2}{N^2} \right) \frac{8 \Tbar^3}{\left( f\mathcal{N} - \frac{{N}-4}{{N}} \right) \left( f\mathcal{N} - \frac{{N}-2}{{N}}  \right)}  \nonumber \\
    &= \lrangleN{\delta T_i^3} \frac{1}{\left( 1 - \frac{{N}-4}{{N}} \frac{1}{f\mathcal{N}}\right) \left( 1 - \frac{{N}-2}{{N}}\frac{1}{f\mathcal{N}}  \right)}
    \approx \lrangleN{\delta T_i^3} \left(1 + \left(2-\frac{6}{N}\right) \frac{1}{f\mathcal{N}} 
    \right) \, ,
    \\
    \lrangleM{\delta T_i^2 \delta T_j} 
    &= - \frac{N-2}{N^2} \frac{8 \Tbar^3}{\left( f\mathcal{N} - \frac{{N}-4}{{N}} \right) \left( f\mathcal{N} - \frac{{N}-2}{{N}}  \right)}
    \approx \lrangleN{\delta T_i^2\delta T_j} \left(1 + \left(2-\frac{6}{N}\right) \frac{1}{f\mathcal{N}} 
    \right) \, ,
    \\
    \lrangleM{\delta T_i \delta T_j \delta T_k} 
    &= \frac{2}{N^2} \frac{8 \Tbar^3}{\left( f\mathcal{N} - \frac{{N}-4}{{N}} \right) \left( f\mathcal{N} - \frac{{N}-2}{{N}}  \right)}
    \approx \lrangleN{\delta T_i \delta T_j \delta T_k} \left(1 + \left(2-\frac{6}{N}\right) \frac{1}{f\mathcal{N}} 
    \right) \, .
\end{align}

The expressions above incorporate the conservation corrections derived in Section~\ref{sec:ConservationCorrection} together with an additional correction due to the discretized multiplicative-noise dynamics. 
The conservation corrections depend only on $N$ and arise from the physical constraint of energy conservation in an isolated system. 
By contrast, the remaining factors proportional to $1/(f\mathcal{N})$ represent modeling errors introduced by the finite-volume approximation. 
These errors become negligible in the limit of large $f\mathcal{N}$, in which case all of the moments and correlations derived above reduce to their corresponding conservation-corrected expressions from Section~\ref{sec:ConservationCorrection}.
For fixed $f\mathcal{N}$, the coefficients multiplying $1/(f\mathcal{N})$ approach finite limits as $N\rightarrow\infty$. 
Thus, unlike the conservation correction, the modeling error does not vanish with increasing system size and is instead controlled primarily by the number of particles per cell.

\begin{figure*}[htbp]
    \centering
    \includegraphics[width=\textwidth]{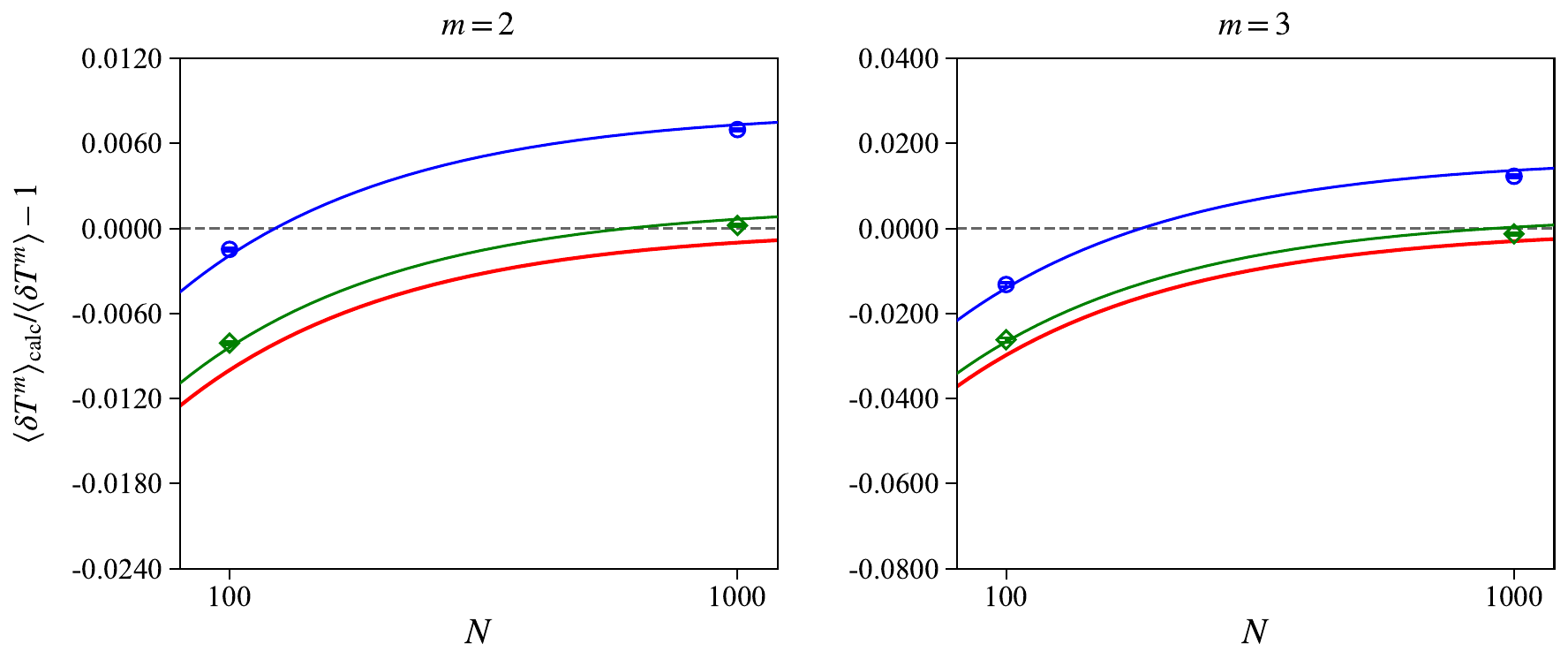}\\
    \caption{Fractional differences for central moments in periodic systems. Left and right panels are $m=2$ and $m=3$, respectively. The blue circles and green diamonds (with error bars) are from simulation measurements of $\langle \delta T^m \rangle_s$ for $\mathcal{N} = 20$ and $\mathcal{N} = 100$ respectively. The blue and green lines, corresponding to $\mathcal{N} = 20$ and $\mathcal{N} = 100$ respectively, are the analytical model predictions. The red line is for the conservation correction $\langle \delta T^m \rangle_{(N)}$.}
    \label{fig:central_moments}
\end{figure*}

Finally, we evaluate how well the theoretical modeling error discussed above predicts the behavior of periodic systems. We consider four systems with $N$ = 100 or 1000 and $\mathcal{N}$ = 20 or 100.  The comparison of simulation results to theory is shown in Figure \ref{fig:central_moments}.  Because the systems are isolated, to obtain the physically correct answer we need to include the finite-size correction, which is shown in red in the Figure. Simulation results for $\mathcal{N}=20$ and $\mathcal{N}=100$ are shown by the symbols in the Figure while the theoretical values are given by the curves.
Overall, there is good agreement between the measured values $\lrangleS{ \delta T^m}$ and the predicted values $\lrangleM{ \delta T^m}$. 
As expected, the modeling error is reduced for larger $\mathcal{N}$.   

\section{Higher-Order Schemes}\label{sec:HigherOrderSchemes}

The analysis in the sections above has been restricted to the Euler--Maruyama scheme.
In this section we examine the model error in the central moments for a higher-order scheme, namely Heun's method or predictor-corrector.
With this temporal integrator, Eq.~\eqref{eq:fvdisc} is replaced by
\begin{subequations}
\begin{align}
    T_i^* 
    &= T_i^n + \frac{\Delta t}{\Delta x} \left( F_{i+\half}^n - F_{i-\half}^n \right) \, , 
    \\
    T_i^{n+1} 
    &= \frac{1}{2} \left\{
        T_i^n + T_i^* + \frac{\Delta t}{\Delta x} \left( F_{i+\half}^* - F_{i-\half}^* \right) 
    \right\} \, ,
\end{align}
\label{eq:fvdiscHeun}
\end{subequations}
where
\begin{equation}
    F_{i+\half}^* = 
    \kappa \frac{T_{i+1}^*-T_i^*}{\Delta x} + \noisecoeff \, T_{i+\half}^* \tilde{Z}_{i+\half}^n \, . 
\end{equation}
Heun's method solves the corresponding stochastic ordinary equation in the Stratonovich interpretation, while Euler--Maruyama gives the It\^{o} interpretation for the stochastic calculus~\cite{kloeden2012numerical}.

As in Section~\ref{sec:ModelErrorsDirichlet}, we consider the one-cell case with a Dirichlet boundary condition on the left and a homogeneous Neumann boundary condition on the right.
For Heun's method, Eq.~\eqref{eq:OneCellFE} is replaced by
\begin{subequations}
\begin{align}
    T^* 
    &= T + 2 A ( T_\mathrm{L} - T ) - \sqrt{2} B\, \mathcal{D}_\zeta(T_\mathrm{L}, T)\, Z \, ,  
    \\
    T' 
    &= \frac12 \left\{
        T + T^* + 2A(T_\mathrm{L} - T^*) - \sqrt{2} B\, \mathcal{D}_\zeta(T_\mathrm{L}, T^*)\, Z 
    \right\} \, .
\end{align}
\label{eq:OneCellPC}
\end{subequations}
Note that the predictor and corrector steps use the same value for $Z$.
One can verify that, in the limit as $\Delta t \rightarrow 0$,
the moments resulting from Eq.~\eqref{eq:OneCellPC} are the same as those obtained from the Euler--Maruyama  discretization with an additional term given by the It\^{o}/Stratonovich correction.
Following the procedure described in Section~\ref{sec:ModelErrorsDirichlet} allows us to obtain the central moments.
However, it is important to note that, due to the change in the interpretation of stochastic calculus, for $\zeta = \half$ the mean value of the temperature is 
\begin{equation}
    \lrangleM{T} = \mathcal{S} T_\mathrm{L} \, ,
    \quad \mathrm{where} \quad
    \mathcal{S} = \left( 1 + \frac{1}{f\mathcal{N} - \frac12}\right) \, ,
\label{eq:HeunDirichletShift}
\end{equation}
so we set $T_\mathrm{L} = \overline{T}/\mathcal{S}$ to have $\lrangleM{ \delta T } = 0$.

For $\zeta = 0$, the noise is additive and the limiting behavior as $\Delta t \rightarrow 0$ is the same as for Euler--Maruyama.
For $\zeta = \frac12$ the variance is
\begin{equation}
    \lrangleM{\delta T^2} 
    = \lrangle{\delta T^2} \frac{1}{1 - \frac{3}{4}\frac{1}{(f\mathcal{N})^2}-\frac{1}{4}\frac{1}{(f\mathcal{N})^3}} 
    \approx \lrangle{\delta T^2} \left(1 + \frac{3}{4}\frac{1}{(f\mathcal{N})^2}\right) \, .
\label{eq:OneCellMom2Heun}
\end{equation}
Comparing this with Eq.~\eqref{eq:OneCellMom2EM}, we see that with Heun's method (Stratonovich calculus) the modeling error in the variance is higher order in $1/(f\mathcal{N})$.
The third and fourth central moments are
\begin{align}
    \lrangleM{\delta T^3} 
    &= \lrangle{\delta T^3} \frac{1}{1 - \frac{1}{\FN} -\frac{3}{2}\frac{1}{(\FN)^2}
    +\frac{1}{2}\frac{1}{(\FN)^3}+\frac{13}{16}\frac{1}{(\FN)^4}+\frac{3}{16}\frac{1}{(\FN)^5}}
    \approx \lrangle{\delta T^3} \left(1 + \frac{1}{f\mathcal{N}}\right) \, ,
\label{eq:OneCellMom3Heun}
    \\
    \lrangleM{\delta T^4} 
    &= \lrangle{\delta T^4} \frac{1+\frac{5}{2\FN}}{1+\frac{3}{2\FN}-\frac{11}{(\FN)^2}
    -\frac{1}{4(\FN)^3}+\frac{265}{16(\FN)^4}+\frac{151}{32(\FN)^5}-\frac{231}{32(\FN)^6}-\frac{73}{16(\FN)^7}-\frac{3}{4(\FN)^8}} \nonumber \\
    &\approx \lrangle{\delta T^4} \left(1 + \frac{1}{f\mathcal{N}}\right) \, .
\label{eq:OneCellMom4Heun}
\end{align}
Comparing Eq.~\eqref{eq:OneCellMom3Heun} with Eq.~\eqref{eq:OneCellMom3EM} and  Eq.~\eqref{eq:OneCellMom4Heun} with Eq.~\eqref{eq:OneCellMom4EM}, we see that the Heun method gives similar $O(1/\FN)$ errors with slightly smaller coefficients than the Euler--Maruyama discretization.

A complete analysis of the modeling errors for a system with $N$ cells is straightforward but extremely tedious. 
Yet the results obtained for the single-cell system illustrate two important points.
First, higher-order schemes, such as Heun's method, can introduce a correction to the Dirichlet boundary condition, such as in Eq.~\eqref{eq:HeunDirichletShift}.
Second, the model error in the central moments can be reduced but not eliminated by a higher-order method.

\section{Compressible Navier--Stokes Equations}

\subsection{Formulation}

A natural question to ask is whether fluctuating hydrodynamic models can capture higher-order moments in more complex systems.
Specifically, we consider the stochastic equations for a single component, compressible fluid with multiplicative noise, written as~\cite{srivastava2023staggered}:
\begin{subequations}
\begin{align}
    &\frac{\partial}{\partial t} \left( \rho \right) = - \bnabla\bcdot\left( \rho \uFluid \right) \, ,
\label{eq:CompressibleFHD_Rho}
    \\
    &\frac{\partial }{\partial t} \left( \rho \uFluid \right) =  - \bnabla \bcdot \left[ \rho \uFluid \otimes \uFluid + p\mathbb{I} \right] - \bnabla \bcdot \left[ \overline{\StressTensor} + \widetilde{\StressTensor} \right] \, ,
\label{eq:CompressibleFHD_Mom}
\\                      
    &\frac{\partial }{\partial t} \left( \rho \ESpecific \right) = - \bnabla \bcdot \left[\uFluid \left(\rho \ESpecific + p\right) \right] - \bnabla \bcdot \left[ \overline{\HeatFlux} + \widetilde{\HeatFlux} \right] -\bnabla \bcdot \left[ \left( \overline{\StressTensor} + \widetilde{\StressTensor} \right) \bcdot  \uFluid \right] \, ,
\label{eq:CompressibleFHD_E}
\end{align}
\label{eq:CompressibleFHD}
\end{subequations}
where $\rho$ is the fluid density, $\uFluid$ is the velocity, $\ESpecific$ is the total specific energy, $p$ is the pressure, and $\mathbb{I}$ is the identity tensor. 
The total energy density of the fluid $\rho \ESpecific = \rho e + \frac12 \rho (\uFluid \bcdot \uFluid)$ is the sum of internal energy and kinetic energy, where $e = c_V T$ is the specific internal energy. 
We assume an ideal gas equation of state,
$p = {\rho \kB T}/{m}$, and a constant specific heat, which for a classical monatomic gas ($f = 3$) is $c_V = \frac32 \kB / m$.

In the nonlinear fluctuating hydrodynamic equations, the stress tensor and heat flux have deterministic, $\overline{\StressTensor}$ and $\overline{\HeatFlux}$, plus stochastic, $\widetilde{\StressTensor}$ and $\widetilde{\HeatFlux}$, contributions.
When $\widetilde{\StressTensor}=\widetilde{\HeatFlux}=0$, Eq.~\eqref{eq:CompressibleFHD} reduces to the \emph{deterministic} Navier--Stokes equations for compressible fluids.
For constant $\rho$ and $\uFluid = 0$, Eq.~\eqref{eq:CompressibleFHD_E} can be rewritten as the stochastic heat equation.

The components of the deterministic stress tensor $\StressTensor$ defined in its Newtonian form are:
\begin{equation}
    \overline{\StressTensor}_{ij} 
    = -\eta \left( \frac{\partial u_i}{\partial x_j} + \frac{\partial u_j}{\partial x_i} \right) 
    - \delta_{ij} \left( ( \eta_B - \frac{2}{3} \eta ) {\bf \nabla} \cdot {\uFluid} \right) \, ,
\label{eqn:stress}
\end{equation}
where $\delta_{ij}$ is the Kronecker delta, $\eta$ is the shear viscosity, and $\eta_B$ is the bulk viscosity. 
The deterministic heat flux is $\overline{\HeatFlux} = -\lambda \nabla T$, where $\lambda$ is the thermal conductivity. 

The stochastic stress $\widetilde{\StressTensor}$ is a Gaussian random field 
that is white in space and time with a specific local covariance.
Here, we use the following efficient form of $\widetilde{\StressTensor}$ proposed by~\cite{LLNS_Espanol,morozov1984langevin}:
\begin{equation}
    \widetilde{\StressTensor}(\boldsymbol{r},t) 
    = \sqrt{2\kB  \eta T} \,\widetilde{\boldsymbol{\mathcal{Z}}} 
    + \left(\sqrt{\frac{\kB \eta_B T}{3}} - \frac{\sqrt{2\kB \eta T}}{3} \right) 
    \mathrm{Tr} ( \widetilde{\boldsymbol{\mathcal{Z}}} ) \mathbb{I} \, .
\label{eq:stochstress}
\end{equation}
Here,
\begin{equation}
    \widetilde{\boldsymbol{\mathcal{Z}}} = \frac{1}{\sqrt{2}}\left(\boldsymbol{\mathcal{Z}}+\boldsymbol{\mathcal{Z}}^{T}\right) 
    \label{stoch:Z}
\end{equation}
is a symmetric matrix constructed from an uncorrelated Gaussian tensor field $\boldsymbol{\mathcal{Z}}$ with zero mean and unit variance. 
The stochastic heat flux $\widetilde{\HeatFlux}$ is given by
\begin{equation}
    \widetilde{\HeatFlux} = \sqrt{2\kB \lambda T^2}\boldsymbol{\mathcal{Z}}^{(\HeatFlux)} \, ,
\end{equation}
where $\mathcal{Z}^{(\HeatFlux)}$ is an uncorrelated three-dimensional Gaussian vector field with zero mean and unit variance.

\subsection{Numerics and Simulation Results}

A staggered-grid discretization based on the method-of-lines approach is used to spatially discretize the stochastic PDEs of compressible FHD. 
Here, the conserved scalar variables, $\rho$ and $\rho E$, and primitive scalar variables, $p$ and $T$, are discretized at the centers of a finite-volume cell, whereas the vector variables, conserved momentum density $\rho \uFluid$ and velocity $\uFluid$, are discretized on the normal faces of the grid~\cite{srivastava2023staggered}. 
The resulting stochastic ordinary differential equations (ODEs) are integrated explicitly in time using a low-storage third-order Runge--Kutta (RK3) integrator~\cite{Donev_CAMCOS_10,srivastava2023staggered}. 
The staggered-grid numerical method discretely preserves the fluctuation-dissipation balance~\cite{balboa2012staggered}, which has been confirmed by a correct reproduction of the structure factors of hydrodynamic variables at thermodynamic equilibrium~\cite{srivastava2023staggered}.
Here we only consider closed systems with either periodic boundary conditions or rigid, stationary walls; for the latter case the mass flux is strictly zero at the boundaries.

The simulation parameters are: $\overline{T} = 300~\mathrm{K}$; 
$\overline{\rho} = 0.817~\mathrm{kg}/\mathrm{m}^{3}$; 
$\Delta x = 2.017\times10^{-8}~\mathrm{m}$; $\mathcal{A} = 4.068\times10^{-16}~\mathrm{m}^2$, $N = 100$.
The viscosity and thermal conductivity  are those of a dilute neon gas; they are not treated as constants but depend on the local state of the fluid~\cite{Giov1}.
The average number of particles per grid cell is $\lrangle{\mathcal{N}} \approx 200$ so the temperature fluctuations are comparable to those of the SHE for $\mathcal{N} = 100$ since $f = 3$ for a monatomic ideal gas.

\begin{figure}[htbp]
    \centering
    \includegraphics[width= \textwidth]{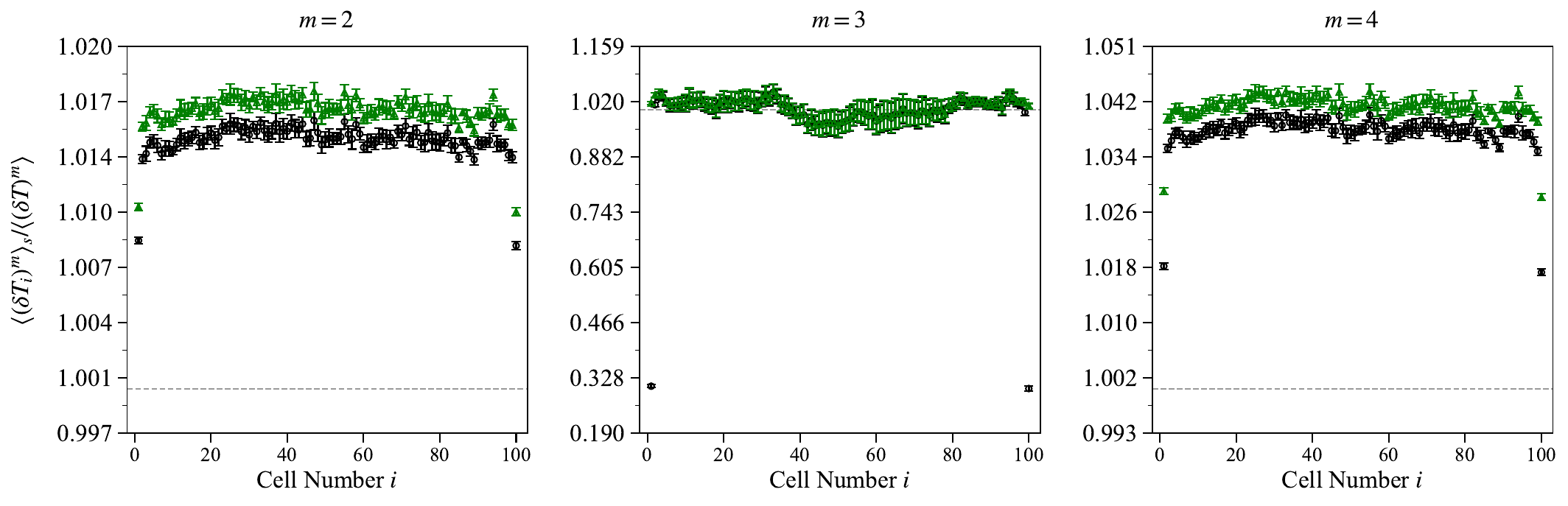}
    \caption{Measured normalized central moments $\lrangleS{ \delta T^m_i } / \lrangle{ \delta T^m_i}$ versus cell index $i$ for $m=2$ (left), $m=3$ (center), and $m=4$ (right). Systems have $N = 100$ cells and Dirichlet boundary conditions with either $\zeta = 0$ (black circles) or $\zeta = \frac12$ (green triangles).}
    \label{fig:MomentsDirichletFHD}
\end{figure}

Figure~\ref{fig:MomentsDirichletFHD} compares the measured normalized central moments, $\lrangleS{ \delta T^m_i } / \lrangle{ \delta T^m_i}$, for the two options for Dirichlet boundary conditions with $T_\mathrm{L} = T_\mathrm{R} = \overline{T}$.
The time step used, $\Delta t = 5 \times 10^{-13}$~s, was found to give accurate results based on comparisons to simulation data for larger values of $\Delta t$.
As with SHE, the traditional choice, $\zeta = 0$, performs well for the second and fourth moments, but has a very large error in the third moment next to the boundary.
By contrast, the other option, $\zeta = \frac12$, gives good results (fractional difference under one percent) for all three moments.

\begin{figure}[htbp]
    \centering
    \includegraphics[width= \textwidth]{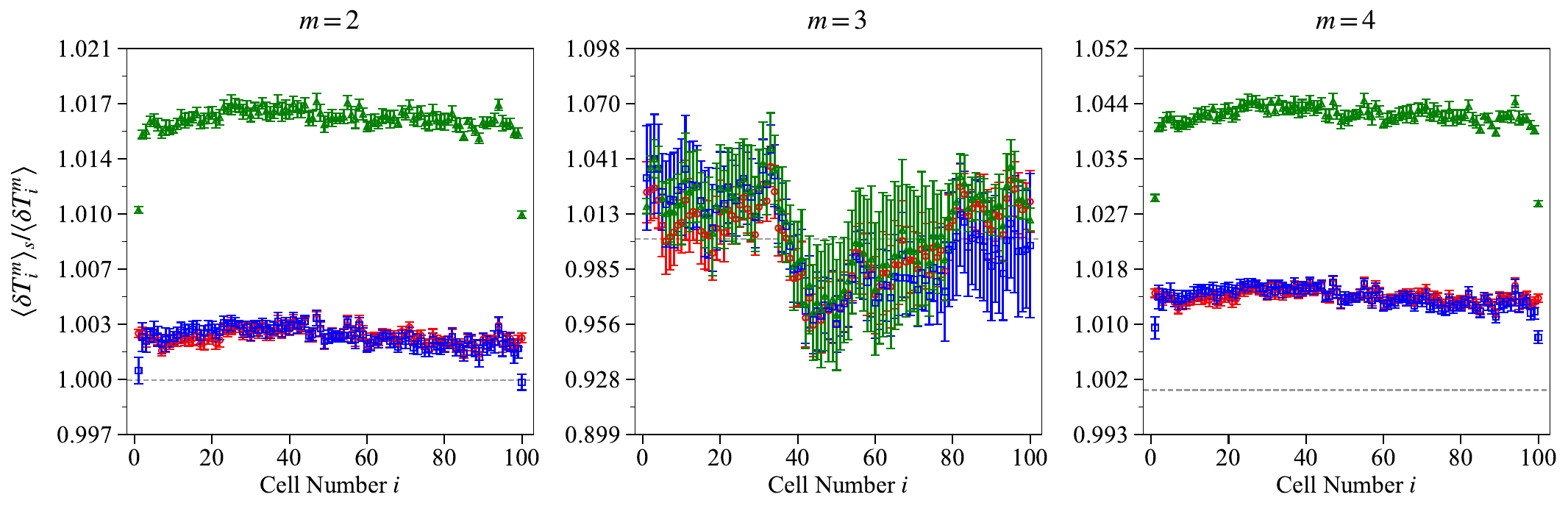}
    \caption{Measured normalized central moments $\lrangleS{ \delta T^m_i } / \lrangle{ \delta T^m_i}$ versus cell index $i$ for $m=2$ (left), $m=3$ (center), and $m=4$ (right). Systems have $N = 100$ cells and either periodic (red circles), Neumann (blue squares), or Dirichlet with $\zeta = \frac12$ (green triangles) boundary conditions.}
    \label{fig:MomentsBCsFHD}
\end{figure}

Figure~\ref{fig:MomentsBCsFHD} shows the measured normalized central moments, $\lrangleS{ \delta T^m_i } / \lrangle{ \delta T^m_i}$,  for systems with various boundary conditions. 
As with SHE for Neumann and Dirichlet boundary conditions, the cells near the boundaries are indistinguishable from cells in the interior.
Due to the conservation correction, which is not included in the plots, the second and fourth moments for periodic and Neumann boundary conditions are smaller than for Dirichlet boundary conditions.
Here the fractional correction due to conservation, $\lrangleN{ \delta T_i^m }/\lrangle{ \delta T_i^m }$, equals 0.9900, 0.9702, and 0.9798 for $m = 2, 3$ and 4, respectively.
Note that in all cases the error bars for the fourth moment are relatively small since its value is dominated by the non-excess kurtosis (i.e., by the $3 \lrangle{\delta T^2}^2$ term).

\begin{figure}[htbp]
    \centering
    \includegraphics[width=\textwidth]{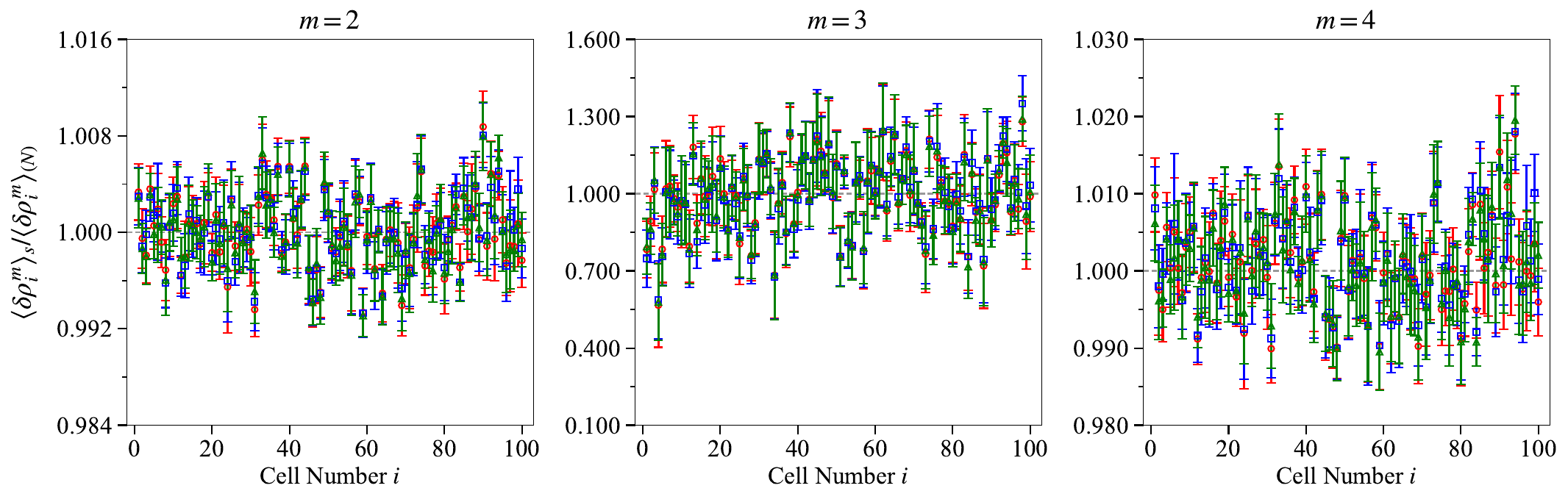}
    \caption{Measured normalized central moments $\lrangleS{ \delta \rho^m_i } / \lrangleN{ \delta \rho^m_i}$ versus cell index $i$ for $m=2$ (left), $m=3$ (center), and $m=4$ (right). Systems have $N = 100$ cells and either periodic (red circles), Neumann (blue squares), or Dirichlet (green triangles) boundary conditions.    }
    \label{fig:DensityMoments}
\end{figure}

Finally, we consider the central moments for density fluctuations.
For an ideal gas, the probability distribution for the number of particles in a cell, $\mathcal{N}$, is the Poisson distribution; the first few central moments are
\begin{equation}
    \lrangle{\delta \mathcal{N}^2} = \lrangle{\mathcal{N}} \, ,
    \quad
    \lrangle{\delta \mathcal{N}^3} = \lrangle{\mathcal{N}} \, ,
    \quad
    \lrangle{\delta \mathcal{N}^4} = 3 \lrangle{\mathcal{N}}^2 + \lrangle{\mathcal{N}} \, .
\end{equation}
For the mass density, $\rho = m \mathcal{N}/V$, the central moments are $\langle \delta \rho^n \rangle = ( m / V )^n \lrangle{ \delta \mathcal{N}^n }$.
Since mass is conserved in our closed systems there are conservation corrections for these central moments.
From the results in Section~\ref{sec:ConservationCorrection} we have for $N$ cells
\begin{align}
    \lrangleN{\delta \rho_i^2} 
    &=  \left(1 - \frac1N\right) \lrangle{\delta \rho^2} \, ,
    \\
    \lrangleN{\delta \rho_i^3} 
    &=  \left(1 - \frac3N + \frac{2}{N^2}\right) \lrangle{\delta \rho^3} \, ,
    \\
    \lrangleN{ \delta \rho_i^4 } 
    &= \left(3 + \exkurt - \frac{6 + 4 \exkurt}{N} +\frac{3 + 6 \exkurt}{N^2} - \frac{3 \exkurt}{N^3} \right) \lrangle{\delta \rho^2}^2 \, ,
\end{align}
where $\exkurt = 1/\mathcal{N}$ is the excess kurtosis for the Poisson distribution.
Figure~\ref{fig:DensityMoments} shows the measured normalized central moments, $\lrangleS{ \delta \rho^m_i } / \lrangleN{ \delta \rho^m_i}$,  for systems with various boundary conditions.
In all cases, the simulation results agree well with the theoretical values after accounting for the conservation corrections.

In summary, the general observations regarding the central moments of temperature in the stochastic heat equation also apply to the full, compressible equations of fluctuating hydrodynamics.
The only significant difference is that for the latter the conservation correction, while present for mass and energy density, does not appear for temperature fluctuations~\cite{garcia2007estimating}.

\section{Summary and Conclusions}

In this work, we investigated the evaluation of central moments up to fourth order in computational fluctuating hydrodynamics (CFHD) using finite-volume discretizations. 
While traditional validation of CFHD algorithms focuses primarily on second-order central moments (i.e., variance), accurately resolving higher-order moments such as skewness and kurtosis is essential for predicting nonlinear functions of fluctuating fields, including nucleation rates and chemical reaction equilibrium constants. 
By analyzing the one-dimensional stochastic heat equation, we demonstrated that incorporating multiplicative noise is necessary to obtain the gamma distribution for temperature fluctuations, as expected from the canonical ensemble, while additive noise incorrectly yields a Gaussian distribution with zero skewness. 
Furthermore, we derived analytical expressions for conservation-induced finite-size corrections of $O(1/N)$ in isolated systems, as well as modeling errors of $O(1/(f\mathcal{N}))$ arising from the discrete representation of the systems as a system of stochastic ODEs.
We also established that an alternative formulation for Dirichlet boundary conditions ($T_0 = \frac12 (T_\mathrm{L} + T_1)$) correctly captures higher-order moments, resolving severe skewness discrepancies present in the traditional finite-volume boundary treatment ($T_0 = T_\mathrm{L}$).
Given the importance of the multiplicative nature of the stochastic noise, we also investigated higher-order temporal integrators, showing that Stratonovich-based schemes such as the Heun predictor-corrector method reduce the leading modeling error in variance to $O(1/(f\mathcal{N})^2)$. 
Finally, numerical measurements from fluctuating hydrodynamic simulations of the full compressible, stochastic Navier--Stokes equations confirmed the practical applicability and generality of the results obtained for the stochastic heat equation. 

The present work lends itself to the study of other fluctuating hydrodynamics equations.
For example, the Dean--Kawasaki equation~\cite{dean1996langevin,kawasaki1994stochastic,illien2025dean}, which models the diffusion of random walkers in one spatial dimension, is
\begin{equation}
    \partial_t\, \rho =  D\, \partial_x^2 \rho + \partial_x \sqrt{2 D \rho}~\widetilde{Z} \, .
    \label{eq:DK}
\end{equation}
Here, $\rho$ is the number density and $D$ is the diffusion constant.
There is a close similarity between Eq.~\eqref{eq:DK} and the stochastic heat equation, with the significant difference being the non-linear multiplicative noise in the Dean--Kawasaki equation.
There are other related equations, for example adding an advection term to this equation gives the stochastic Burgers equation~\cite{Bell_JCP_07}.

Augmenting the Dean--Kawasaki equation to $N_s$ species diffusing and undergoing $N_r$ reactions gives the reaction-diffusion equation.
Modeling the stochastic reactions by the chemical Langevin equation, we write this as
\begin{equation}
    \partial_t\, \rho_s 
    = D_s \partial_x^2 \rho_s 
    + \partial_x \sqrt{2 D_s \rho_s} ~Z_s^{(D)}
    + \sum_{r=1}^{N_r}\nu_{sr}\left(a_r(\boldsymbol{\rho}) 
    + \sqrt{a_r(\boldsymbol{\rho})} ~Z_r^{(R)}\right) \, ,
\end{equation}
where $D_s$ is the diffusion coefficient of species $s$, $\boldsymbol{\rho} = 
(\rho_1, \ldots, \rho_{N_s})$, $a_r(\boldsymbol{\rho})$ is the propensity function indicating the rate of reaction $r$, and $\nu_{sr}$ is the stoichiometric coefficient of species $s$ in reaction $r$.
Given that the propensity function is typically a non-linear function of number density, accurate numerical solutions of the reaction-diffusion equation require capturing the moments of $\boldsymbol{\rho}$ correctly.

The accuracy of numerical fluctuating hydrodynamics (FHD) schemes is typically assessed by their ability to reproduce the correct equilibrium structure factor. 
The present findings on higher-order moments and correlations indicate that three-point and four-point structure factors, commonly measured in solids, offer a promising new avenue for such validation. 
Future work can build upon these results by conducting formal spatial and temporal truncation error analyses for schemes incorporating multiplicative noise. 
Of particular interest are investigations into alternative spatial discretization strategies (e.g., cell-centered versus staggered grids, positivity preserving flux-limiters~\cite{magaletti2022positivity}) and advanced temporal integration schemes to mitigate modeling errors. 
Finally, while this study focused on finite-volume methods, computational fluctuating hydrodynamics encompasses other numerical frameworks, such as finite element~\cite{de2015finite,martinez2024finite,gourzoulidis2026finite} and spectral methods~\cite{cartes2022galerkin,gosteva2026emergent}. 
Evaluating these alternative schemes using the metrics introduced here represents an important direction for future research.


\begin{acknowledgments}
This material is based upon work supported by the U.S.\ Department of Energy (DOE), Office of Science, Office of Advanced Scientific Computing Research (ASCR), Funding for Accelerated, Inclusive Research (FAIR) program under Award Number DE-SC0024240 and under Contract Number DE-AC02-05CH11231.
The work was also supported by the U.S.\ Department of Energy, Office of Science, Office of Advanced Scientific Computing Research, Applied Mathematics Program under Contract Number DE-AC02-05CH11231.
\end{acknowledgments}

\appendix

\section{Canonical Ensemble and Gamma Distribution}\label{app:GammaDist}

Here we show that the probability distribution for temperature in the canonical ensemble is the gamma distribution.
Starting with the recurrence relation for energy:
\begin{equation}
    \langle({E}-{U})^{n+1}\rangle 
    = -\frac{\partial}{\partial\beta}\langle({E}-{U})^n\rangle 
    - n\langle({E}-{U})^{n-1}\rangle\frac{\partial U}{\partial\beta} \, ,    
\end{equation}
where $\beta = 1 / (\kB\Tbar)$.
We also define
\begin{equation}
    \alpha = \frac{\rho c_V V}{\kB} \, .    
\end{equation}
We then have
\begin{equation}
    U = \frac{\alpha}{\beta} \, ,
    \quad
    \langle(E-U)^{2}\rangle = \frac{\alpha}{\beta^2} \, .
\label{eq:recur_init}
\end{equation}
If we let $\mu = E - U$ then the recurrence relation can be written as
\begin{equation}
    \langle\mu^{n+1}\rangle = 
    -\frac{\partial}{\partial\beta}\langle\mu^n\rangle 
    -n\langle\mu^{n-1}\rangle\frac{\partial{U}}{\partial\beta} \, .
\label{eq:recur_mu}
\end{equation}
Noting that
\begin{equation}
    \frac{\partial U}{\partial\beta} = - \frac{\alpha}{\beta^2} \, ,
\end{equation}
it follows from Eqs.~\eqref{eq:recur_init} and \eqref{eq:recur_mu} that $\langle\mu^{n}\rangle \propto \beta^{-n}$. 
Using this observation, we can write the recurrence relation as
\begin{equation}
    \langle\mu^{n+1}\rangle 
    = \frac{n}{\beta}\langle\mu^n\rangle 
    + \frac{n \alpha}{\beta^2} \langle\mu^{n-1}\rangle \, .
\label{eq:recur_mu_alt}
\end{equation}

We define $M_\mu$ as the moment generating function
\begin{equation}
    M_\mu (t) = \sum_{n=0}^{\infty} \frac{t^n}{n!} \langle \mu^n \rangle \, .    
\end{equation}
Multiply the recurrence relation by $\frac{t^n}{n!}$ and sum over $n$:
\begin{equation}
    \sum_{n=0}^{\infty} \langle\mu^{n+1}\rangle \frac{t^n}{n!} 
    = \sum_{n=0}^{\infty} \langle\mu^n\rangle \frac{n}{\beta}\frac{t^n}{n!} 
    + \sum_{n=1}^{\infty} \langle\mu^{n-1}\rangle \frac{n \alpha}{\beta^2} \frac{t^n}{n!} \, .
\end{equation}
The left hand side can be written as
\begin{equation}
    \sum_{n=1}^{\infty} \langle\mu^n\rangle \frac{t^{n-1}}{(n-1)!} 
    = \frac{\partial}{\partial t} \sum_{n=0}^{\infty} \langle\mu^n\rangle \frac{t^n}{n!} 
    = \frac{\partial M_\mu}{\partial t} \, .
\end{equation}
The first term on the right hand side can be rewritten as
\begin{equation} 
    \sum_{n=0}^{\infty} \langle\mu^n\rangle \frac{n}{\beta}\frac{t^n}{n!} 
    = \frac{t}{\beta} \sum_{n=1}^{\infty} \langle\mu^n\rangle \frac{t^{n-1}}{(n-1)!} 
    = \frac{t}{\beta} \frac{\partial M_\mu}{\partial t} \, . 
\end{equation}
and the second term on the right hand side can be rewritten as
\begin{equation}
    \sum_{n=1}^{\infty}\langle\mu^{n-1}\rangle \frac{n \alpha}{\beta^2} \frac{t^n}{n!} 
    = \frac{\alpha t}{\beta^2} \sum_{n=1}^{\infty}\langle\mu^{n-1}\rangle \frac{t^{n-1}}{(n-1)!} 
    = \frac{\alpha t}{\beta^2} M_\mu \, .
\end{equation}

Combining these terms, we obtain
\begin{equation}
    \frac{\partial M_\mu}{\partial t} = \frac{\alpha t}{\beta(\beta-t)} M_\mu \, .
\label{eq:mgf_ode}
\end{equation}
Since $M_\mu(0) = 1$, we can integrate Eq.~\eqref{eq:mgf_ode} to obtain
\begin{equation}
    M_\mu(t) = \left(1-\frac{t}{\beta}\right)^{-\alpha} e^{-\frac{\alpha t}{\beta}} \, .    
\end{equation}
Using the relationship between the moments
\begin{equation}
    M_{E}(t)= \langle e^{E t} \rangle = \langle e^{t(E-U)} e^{t U} \rangle  = e^{t U} M_{\mu} \, ,
\end{equation}
we then have
\begin{equation}
    M_{E}(t) = \left(1-\frac{t}{\beta}\right)^{-\alpha} \, ,    
\end{equation}
from which follows that the distribution of $E$ is a gamma distribution with shape  $\rho c_V V / \kB$
and rate $1/(\kB \Tbar)$.  
Given that $T = E / {\rho c_V V}$, we then have that the distribution of $T$ is also a gamma distribution but with rate given by $\GammaRate = \rho c_V V / (\kB \Tbar)$.
Finally, the gamma distribution is uniquely defined by its moments since it satisfies Carleman's condition~\cite{akhiezer2020classical}.

We note that the gamma distributions of $E$ and $T$ can easily be derived from the Einstein model of a solid in the classical (high-temperature) limit.
In this regime, the model has $f=6$ degrees of freedom per atom and its total energy is written as
\begin{equation}
    E = \sum_{i=1}^{\mathcal{N}} \sum_{\alpha\in\{x,y,z\}} \left(
        \frac12 k\, r_{i,\alpha}^2 + \frac12 m\, v_{i,\alpha}^2
        \right) \, ,
\label{eq:EEinstein}
\end{equation}
where $k$ is the force constant and $m$ is the mass of an atom.
Each component of the displacement and velocity of each atom, denoted by $r_{i,\alpha}$ and $v_{i,\alpha}$, is independent and normally distributed:
\begin{equation}
    r_{i,\alpha} \sim \mathrm{Normal}\left(\mu=0,\sigma^2=\frac{\kB \overline{T}}{k}\right) \, ,
    \quad
    v_{i,\alpha} \sim \mathrm{Normal}\left(\mu=0,\sigma^2=\frac{\kB \overline{T}}{m}\right) \, ,
\end{equation}
where $\mu$ and $\sigma^2$ denote the mean and variance parameters of a normal distribution.
Thus, each potential- or kinetic-energy contribution follows a gamma distribution:
\begin{equation}
     \frac12 k\, r_{i,\alpha}^2 \sim 
     \mathrm{Gamma}\left(\alpha = \frac12,\beta=\frac{1}{\kB \overline{T}}\right) \, , 
     \quad
     \frac12 m\, v_{i,\alpha}^2 \sim
     \mathrm{Gamma}\left(\alpha = \frac12,\beta=\frac{1}{\kB \overline{T}}\right) \, ,
\end{equation}
where $\alpha$ and $\beta$ denote the shape and rate parameters of a gamma distribution.
From Eq.~\eqref{eq:EEinstein}, $E$ and thus $T = E / (\rho c_V V) = E / (3 \kB \mathcal{N})$ also follow gamma distributions:
\begin{equation}
    E \sim \mathrm{Gamma}\left(\alpha = 3\mathcal{N},\beta=\frac{1}{\kB \overline{T}}\right) \, , 
    \quad
    T \sim \mathrm{Gamma}\left(\alpha = 3\mathcal{N},\beta=\frac{3\mathcal{N}}{\overline{T}}\right) \, .
\end{equation}
For an ideal monatomic gas, only the kinetic-energy contributions are retained. 
Summing the $3\mathcal{N}$ independent quadratic contributions gives
\begin{equation}
    E \sim \mathrm{Gamma}\left(\alpha = \frac32\mathcal{N},\beta=\frac{1}{\kB \overline{T}}\right) \, , 
    \quad
    T \sim \mathrm{Gamma}\left(\alpha = \frac32\mathcal{N},\beta=\frac{3\mathcal{N}}{2\overline{T}}\right) \, .
\end{equation}

\section{Conservation Correction for Fourth Moment}\label{app:4thMom}

We assume
\begin{equation}
\begin{split}
    \langle \delta T_i \delta T_j \delta T_k \delta T_l \rangle_{(N)} 
    & = a_1 \delta_{ij} \delta_{jk} \delta_{kl}
    + a_2 ( \delta_{ij}\delta_{kl} + \delta_{ik}\delta_{jl} + \delta_{il}\delta_{jk} ) 
    \\
    & + a_3 ( \delta_{ij}\delta_{ik} + \delta_{ij}\delta_{il} + \delta_{ik}\delta_{il} + \delta_{jk}\delta_{jl} ) 
    \\
    & + a_4 ( \delta_{ij} + \delta_{ik} + \delta_{il} + \delta_{jk} + \delta_{jl} + \delta_{kl} ) 
    + a_5 \, .
\end{split}
\end{equation}
Since $\sum_j \delta T_j = 0$, we have
\begin{align}
    &\sum_j \sum_k \sum_l \langle \delta T_i \delta T_j \delta T_k \delta T_l \rangle_{(N)} 
    = a_1 + 3N a_2 + 4N a_3 + 6N^2 a_4 + N^3 a_5= 0 \, ,
    \\
    &\sum_k \sum_l \langle \delta T_i^2 \delta T_k \delta T_l \rangle_{(N)}
    = a_1 + (N+2) a_2 + (2N+2) a_3 + (N^2 + 5N) a_4 + N^2 a_5 = 0 \, ,
    \\
    &\sum_l \langle \delta T_i^3 \delta T_l \rangle_{(N)}
    = a_1 + 3 a_2 + (N+3) a_3 + (3N+3) a_4 + N a_5 = 0 \, ,
\end{align}
and thus
\begin{equation}
    a_3 = -\frac{1}{N} a_1 \, ,
    \quad
    a_4 = \frac{1}{N^2} a_1 - \frac{1}{N} a_2 \, ,
    \quad
    a_5 = -\frac{3}{N^3} a_1 + \frac{3}{N^2} a_2 \, .
\end{equation}
Since in the thermodynamic limit (i.e., $N\rightarrow\infty$)
\begin{align}
    &\langle \delta T_i^4 \rangle_{(N)} 
    = \left(1-\frac{4}{N}+\frac{6}{N^2} - \frac{3}{N^3}\right) a_1
    + \left(3-\frac{6}{N}+\frac{3}{N^2}\right) a_2
    \rightarrow a_1 + 3 a_2 = \langle \delta T^4 \rangle \, ,
    \\
    &\langle \delta T_i^2 \delta T_j^2 \rangle_{(N)}
    = \left(\frac{2}{N^2}-\frac{3}{N^3}\right) a_1
    + \left(1-\frac{2}{N}+\frac{3}{N^2}\right) a_2
    \rightarrow a_2 = \langle \delta T^2 \rangle^2\quad\mbox{for $i\ne j$} \, ,
\end{align}
we have
\begin{align}
    &a_1 = \langle \delta T^4 \rangle - 3 \langle \delta T^2 \rangle^2
    = \frac{48 \Tbar^4}{f^3 \mathcal{N}^3} \, ,
    \\
    &a_2 = \langle \delta T^2 \rangle^2 = \frac{4 \Tbar^4}{f^2 \mathcal{N}^2} \, .
\end{align}
By introducing
\begin{equation}
    \exkurt = \frac{a_1}{a_2} = \frac{12}{f\mathcal{N}} \, ,
\end{equation}
which equals the excess kurtosis, one can express the finite-size effect of $\lrangleN{ \delta T_i \delta T_j \delta T_k \delta T_l }$ using $a_2 = \lrangle{ \delta T^2 }^2$.
For example,
\begin{equation}
    \lrangleN{\delta T_i^4}
    = \left(3+\exkurt -\frac{6+4\exkurt}{N} + \frac{3+6\exkurt}{N^2} - \frac{3\exkurt}{N^3}\right) \lrangle{ \delta T^2 }^2 \, .
\end{equation}

\section{Multiplicative Noise Analysis for Multiple Cells}\label{app:multiplicative_noise}

The multiplicative noise analysis developed for the single-cell case (see Section~\ref{sec:ModelErrorsDirichletSingle}) can be extended to the multi-cell setting.
To this end, we introduce vector notation and make use of the Kronecker product, denoted by $\otimes$.
Specifically, the state variables are represented by the column vector $\delta\boldsymbol{T}=[\delta T_0,\delta T_1,\ldots,\delta T_{N-1}]^T$, and the second-, third-, and fourth-order moments are given as $\boldsymbol{q}_2=\langle \delta\boldsymbol{T} \otimes \delta\boldsymbol{T} \rangle$, $\boldsymbol{q}_3=\langle \delta\boldsymbol{T} \otimes \delta\boldsymbol{T} \otimes \delta\boldsymbol{T} \rangle$, $\boldsymbol{q}_4=\langle \delta\boldsymbol{T} \otimes \delta\boldsymbol{T} \otimes \delta\boldsymbol{T} \otimes \delta\boldsymbol{T} \rangle$, respectively.
For column vectors $\boldsymbol{x}\in\mathbb{R}^n$ and $\boldsymbol{y}\in\mathbb{R}^m$, the Kronecker product $\boldsymbol{x} \otimes \boldsymbol{y} \in\mathbb{R}^{mn}$ is defined by
\begin{equation}
    \boldsymbol{x} \otimes \boldsymbol{y} 
    = \begin{bmatrix}
        x_1 \boldsymbol{y} \\
        \vdots \\
        x_n \boldsymbol{y}
    \end{bmatrix} \, .
\end{equation}
Accordingly, $\boldsymbol{q}_2$, $\boldsymbol{q}_3$, and $\boldsymbol{q}_4$ are column vectors containing $N^2$, $N^3$, and $N^4$ entries, respectively.

For the Euler--Maruyama scheme, the time update of $\delta\boldsymbol{T}$ can be written in the form
\begin{equation}
    \delta\boldsymbol{T}' 
    = \delta\boldsymbol{T} + A_0\Delta t \mathbf{A}_1\delta\boldsymbol{T}
    + B_0\sqrt{\Delta t}\; \boldsymbol{v}(\delta\boldsymbol{T},\boldsymbol{R}) \, .
\end{equation}
Here, $\delta\boldsymbol{T}'\in \mathbb{R}^N$ denotes the updated state vector after a time step of $\Delta t$, and $\boldsymbol{R}\in \mathbb{R}^M$ denotes $M$ independent standard normal random variates that are used to compute stochastic fluxes at cell faces; see Eq.~\eqref{eq:ZmathfrakN}.
For all boundary conditions considered in this paper, including the Dirichlet boundary with $\mathcal{D}_\zeta(\overline{T},T)$ (see Eq.~\eqref{eq:Dzeta}), the multiplicative noise $\boldsymbol{v}(\delta\boldsymbol{T},\boldsymbol{R})\in\mathbb{R}^N$ can be decomposed into a term linear in $\boldsymbol{R}$ and a term bilinear in $\delta\boldsymbol{T}$ and $\boldsymbol{R}$:
\begin{equation}
    \boldsymbol{v}(\delta\boldsymbol{T},\boldsymbol{R})
    = \overline{T}\,\mathbf{B}_1 \boldsymbol{R} 
    + \mathbf{B}_2(\delta\boldsymbol{T}\otimes\boldsymbol{R}) \, .
\end{equation}
We fix the values of $A_0$ and $B_0$ as
\begin{equation}
    A_0 = \frac{\kappa}{\Delta x^2} \, ,
    \quad
    B_0 = \frac{\noisecoeff}{\Delta x \sqrt{\mathcal{A}\Delta x}} \, ,
\end{equation}
so that the matrices $\mathbf{A}_1\in\mathbb{R}^{N\times N}$, $\mathbf{B}_1\in\mathbb{R}^{N\times M}$, and $\mathbf{B}_2\in\mathbb{R}^{N\times MN}$ do not contain any physical parameters, $\Delta x$, or $\Delta t$. 
For given boundary conditions and a given value of $N$, one can determine these matrices.  

We give here the values of $\mathbf{A}_1$, $\mathbf{B}_1$, and $\mathbf{B}_2$ for some example cases.
For the single-cell case considered in Section~\ref{sec:ModelErrorsDirichletSingle}, we have $M=1$ and
\begin{equation}
    \mathbf{A}_1 = -2 \, ,
    \quad
    \mathbf{B}_1 = -\sqrt{2} \, ,
    \quad
    \mathbf{B}_2 = -\sqrt{2}\zeta \, .
\end{equation}
For the double-cell case considered in Section~\ref{sec:ModelErrorsDirichletMulti}, we have $M=2$ and 
\begin{equation}
    \mathbf{A}_1 = \begin{bmatrix}
        -3 & 1 \\
        1 & -1 \\
    \end{bmatrix} \, , 
    \quad
    \mathbf{B}_1 = \begin{bmatrix}
        -\sqrt{2} & 1 \\
        0 & -1
    \end{bmatrix} \, , 
    \quad
    \mathbf{B}_2 = \begin{bmatrix}
        -\sqrt{2}\zeta & \frac12 & 0 & \frac12 \\
        0 & -\frac12 & 0 & -\frac12
    \end{bmatrix} \, .
\end{equation}
For the case of periodic boundary conditions with $N=3$ (see Section~\ref{sec:ModelErrorsPeriodic}), we have $M=3$ and 
\begin{equation}
    \mathbf{A}_1 = \begin{bmatrix}
        -2 & 1 & 1 \\
        1 & -2 & 1 \\
        1 & 1 & -2
    \end{bmatrix} \, , 
    \;\;
    \mathbf{B}_1 = \begin{bmatrix}
        1 & 0 & -1 \\
        -1 & 1 & 0 \\
        0 & -1 & 1
    \end{bmatrix} \, , 
    \;\;
    \mathbf{B}_2 = \begin{bmatrix}
        \frac12 & 0 & -\frac12 & \frac12 & 0 & 0 & 0 & 0 & -\frac12 \\
        -\frac12 & 0 & 0 & -\frac12 & \frac12 & 0 & 0 & \frac12 & 0 \\
        0 & 0 & \frac12 & 0 & -\frac12 & 0 & 0 & -\frac12 & \frac12
    \end{bmatrix} \, .
\end{equation}

We now derive the equation satisfied by $\boldsymbol{q}_2$ in the limit $\Delta t\rightarrow 0$.
By computing $\langle \delta\boldsymbol{T}'\otimes\delta\boldsymbol{T}'\rangle$ and collecting terms up to $O(\Delta t)$, we obtain
\begin{equation}
\label{eq:ysq}
\begin{split}
    \langle \delta\boldsymbol{T}'\otimes\delta\boldsymbol{T}' \rangle
    &= \langle \delta\boldsymbol{T}\otimes\delta\boldsymbol{T} \rangle 
    + B_0\sqrt{\Delta t}\, \Big[\langle\delta\boldsymbol{T}\otimes\boldsymbol{v}\rangle 
    + \langle \boldsymbol{v}\otimes\delta\boldsymbol{T}\rangle\Big] \\
    &+ A_0\Delta t\, \Big[\langle \delta\boldsymbol{T}\otimes (\mathbf{A}_1\delta\boldsymbol{T})\rangle + \langle(\mathbf{A}_1\delta\boldsymbol{T}) \otimes \delta\boldsymbol{T}\rangle\Big]
    + B_0^2\Delta t\, \langle \boldsymbol{v} \otimes \boldsymbol{v} \rangle + o(\Delta t) \, .
\end{split}
\end{equation}
By stationarity, $\langle \delta\boldsymbol{T}'\otimes\delta\boldsymbol{T}' \rangle = \langle \delta\boldsymbol{T}\otimes\delta\boldsymbol{T} \rangle = \boldsymbol{q}_2$.
By analyzing the remaining terms in Eq.~\eqref{eq:ysq}, we show that the $O(\sqrt{\Delta t})$ contributions vanish, while the $O(\Delta t)$ terms can be expressed in terms of $\boldsymbol{q}_2$.
To this end, we introduce the identity matrix $\mathbf{I}_n\in\mathbb{R}^{n\times n}$ and the commutation matrix $\mathbf{S}_{m,n}\in\mathbb{R}^{mn\times mn}$ and use their properties: $\mathbf{I}_n\boldsymbol{x} = \boldsymbol{x}$ and $\boldsymbol{x}\otimes\boldsymbol{y} = \mathbf{S}_{m,n}(\boldsymbol{y}\otimes\boldsymbol{x})$ for $\boldsymbol{x}\in\mathbb{R}^n$ and $\boldsymbol{y}\in\mathbb{R}^m$.
We also use the mixed-product property of the Kronecker product: If $\mathbf{X}$, $\mathbf{Y}$, $\mathbf{Z}$, and $\mathbf{W}$ are matrices of compatible sizes such that the products $\mathbf{X}\mathbf{Z}$ and $\mathbf{Y}\mathbf{W}$ are defined, then $(\mathbf{X}\otimes\mathbf{Y})(\mathbf{Z}\otimes\mathbf{W}) = (\mathbf{X}\mathbf{Z})\otimes(\mathbf{Y}\mathbf{W})$.

We obtain
\begin{align}
    &\delta\boldsymbol{T}\otimes\boldsymbol{v}
    = \overline{T}(\mathbf{I}_N\otimes \mathbf{B}_1)(\delta\boldsymbol{T}\otimes\boldsymbol{R}) 
    + (\mathbf{I}_N\otimes \mathbf{B}_2)(\delta\boldsymbol{T}\otimes\delta\boldsymbol{T}\otimes\boldsymbol{R}) \, ,
    \\
    &\boldsymbol{v}\otimes\delta\boldsymbol{T}
    = \overline{T}(\mathbf{B}_1\otimes\mathbf{I}_N)(\boldsymbol{R}\otimes\delta\boldsymbol{T}) 
    + ((\mathbf{B}_2 \mathbf{S}_{M,N})\otimes\mathbf{I}_N)(\boldsymbol{R}\otimes\delta\boldsymbol{T}\otimes\delta\boldsymbol{T}) \, .
\end{align}
Since $\boldsymbol{R}$ is independent of $\delta\boldsymbol{T}$ and $\langle\boldsymbol{R}\rangle = 0$, we have $\langle\delta\boldsymbol{T}\otimes\boldsymbol{v}\rangle  = \langle\boldsymbol{v}\otimes\delta\boldsymbol{T}\rangle = 0$.
We obtain
\begin{align}
    &\langle \delta\boldsymbol{T}\otimes (\mathbf{A}_1\delta\boldsymbol{T})\rangle
    = (\mathbf{I}_N \otimes \mathbf{A}_1)\boldsymbol{q}_2 \, ,
    \\
    &\langle (\mathbf{A}_1\delta\boldsymbol{T})\otimes\delta\boldsymbol{T} \rangle
    = (\mathbf{A}_1 \otimes \mathbf{I}_N)\boldsymbol{q}_2 \, .
\end{align}
Using $\langle\delta\boldsymbol{T}\rangle = 0$ and $\langle\boldsymbol{R}\otimes\boldsymbol{R}\rangle = \mathrm{vec}(\mathbf{I}_M)$, where $\mathrm{vec}(\mathbf{X})$ denotes the vectorization of the matrix $\mathbf{X}$, we obtain
\begin{equation}
    \langle \boldsymbol{v}\otimes \boldsymbol{v} \rangle
    = \overline{T}^2(\mathbf{B}_1\otimes \mathbf{B}_1)\,\mathrm{vec}(\mathbf{I}_M) + (\mathbf{B}_2\otimes(\mathbf{B}_2 \mathbf{S}_{M,N}))
    \langle \delta\boldsymbol{T}\otimes\mathrm{vec}(\mathbf{I}_M)\otimes\delta\boldsymbol{T}\rangle \, .
\end{equation}
By substituting $\langle \delta\boldsymbol{T}\otimes\mathrm{vec}(\mathbf{I}_M)\otimes\delta\boldsymbol{T}\rangle = (\mathbf{I}_N\otimes\mathrm{vec}(\mathbf{I}_M)\otimes\mathbf{I}_N)\,\boldsymbol{q}_2$ and all results obtained above into Eq.~\eqref{eq:ysq}, we finally obtain
\begin{equation}
\label{eq:q2}
    \Big\{
        A_0\big(\mathbf{I}_N \otimes \mathbf{A}_1 + \mathbf{A}_1 \otimes \mathbf{I}_N\big)
        + B_0^2\big(\mathbf{B}_2\otimes(\mathbf{B}_2 \mathbf{S}_{M,N})\big) (\mathbf{I}_N\otimes\mathrm{vec}(\mathbf{I}_M)\otimes \mathbf{I}_N)
    \Big\} \boldsymbol{q}_2
    + B_0^2\overline{T}^2 (\mathbf{B}_1\otimes \mathbf{B}_1)\mathrm{vec}(\mathbf{I}_M) = 0 \, .
\end{equation}

By solving Eq.~\eqref{eq:q2} for $\boldsymbol{q}_2$, one can compute the second-order moments.
For example, since the matrices $\mathbf{I}_N$ and $\mathbf{S}_{M,N}$ reduce to unity for $N=M=1$, one can easily obtain the results of $\langle \delta T^2\rangle$ for the single-cell case given in Section~\ref{sec:ModelErrorsDirichletSingle}.
For small values of $N>1$, one can solve Eq.~\eqref{eq:q2} for $\boldsymbol{q}_2$ analytically with the aid of symbolic manipulators.
While the computation of $\boldsymbol{q}_3$ and $\boldsymbol{q}_4$ is more involved, the same procedure can be applied and analytic solutions are still tractable with the aid of symbolic manipulators for small values of $N>1$.



%


\end{document}